\documentclass[onecolumn, 12pt]{IEEEtran}
\usepackage{amsmath,graphicx}
\usepackage{color}%
\usepackage{float}
\usepackage{amsmath,bm,amssymb,amsfonts,amsthm,epsfig}
\usepackage{graphicx}
\usepackage{psfrag}
\usepackage{comment}
\usepackage{url}
\usepackage{cite} 
\usepackage{epstopdf}
\usepackage{lipsum}
\usepackage{tabularx}
\usepackage[ruled]{algorithm2e}

\makeatletter
\newcommand*{\rom}[1]{\expandafter\@slowromancap\romannumeral #1@}
\makeatother

\newtheorem{mydef}{Definition}%
\newtheorem{myrem}{Remark}%

\def\vec{{\rm vec}}

\title{Self Calibration of Scalar and Vector Sensor Arrays}
\author{Krishnaprasad Nambur Ramamohan,~\IEEEmembership{Student Member,} Sundeep Prabhakar Chepuri,~\IEEEmembership{Member,~IEEE,} 
	Daniel Fernandez Comesana, 
	and Geert~Leus,~\IEEEmembership{Fellow,~IEEE} 
	\thanks{Krishnaprasad Nambur Ramamohan is with Delft University of Technology, The Netherlands and Microflown Technologies, The Netherlands (e-mail: k.namburramamohan@tudelft.nl, ramamohan@microflown.com). Sundeep Prabhakar Chepuri is with the Department of Electrical Communications Engineering, Indian Institute of Science, Bangalore 560012, India (e-mail: spchepuri@iisc.ac.in).  Daniel Fernandez Comesana is with Microflown  Technologies, Tivolilaan 205, 6824 BV Arnhem, The Netherlands (e-mail: fernandez@microflown.nl). Geert Leus is with the Faculty of Electrical Engineering, Mathematics and Computer Science, Delft University of Technology, Delft 2628 CD, The Netherlands (e-mail: g.j.t.leus@tudelft.nl). This work was supported in part by the Pratiksha Trust Fellowship, IISc and is part of the ASPIRE project (project 14926 within the STW OTP), which is financed by the Netherlands organization for Scientific Research(NWO).
	}
}
	
\begin{document}
	\setlength\belowcaptionskip{-15ex}
	\maketitle
	\begin{abstract}
	In this work, we consider the problem of joint calibration and direction-of-arrival (DOA) estimation using sensor arrays. This joint estimation problem is referred to as self calibration. Unlike many previous iterative approaches, we propose geometry independent convex optimization algorithms for jointly estimating the sensor gain and phase errors as well as the source DOAs. We derive these algorithms based on both the conventional element-space data model and the covariance data model. We focus on sparse and regular arrays formed using scalar sensors as well as vector sensors. The developed algorithms are obtained by transforming the underlying bilinear calibration model into a linear model, and subsequently by using standard convex relaxation techniques to estimate the unknown parameters. Prior to the algorithm discussion, we also derive identifiability conditions for the existence of a unique solution to the self calibration problem. To demonstrate the effectiveness of the developed techniques, numerical experiments and comparisons to the state-of-the-art methods are provided. Finally, the results from an experiment that was performed in an anechoic chamber using an acoustic vector sensor array are presented to demonstrate the usefulness of the proposed self calibration techniques.
\end{abstract}
	%
\section{Introduction}
	
	The problem of estimating the direction-of-arrival (DOA) of multiple far-field events impinging on an array of spatially distributed sensors has received considerable interest in various fields including communications, radio astronomy, acoustics, and seismology, to list a few. For instance in acoustics, microphone or acoustic pressure sensor (APS) arrays are used for direction estimation. Many advanced algorithms such as minimum variance distortionless response~(MVDR)~\cite{capon1969high}, 
	and multiple signal classification~(MUSIC)~\cite{opac-b1105852} have been developed for DOA estimation. Sparse recovery techniques that leverage the sparse structure of the angular domain whenever only a few sources are present have also been widely used for DOA estimation~\cite{malioutov2005sparse}. These traditional algorithms require more physical sensors than the number of sources for DOA estimation and use the data acquired in the \emph{element-space} domain (i.e., at the output of the antenna elements) or in the covariance (or \emph{co-array}) domain obtained by computing the spatial correlation across the antenna elements.
		
	To reduce sensing and data processing costs, sparse sensing methods are gaining attention~\cite{chepuri2016sparse}. One can resolve and estimate DOAs of as many as $\mathcal{O}(M^2)$ sources using only $M$ physical elements by smartly and irregularly placing the sensor elements in the array such that its co-array domain contains the same information as that of the co-array of a uniform array. Such sensor placements are generally referred to as sparse arrays~\cite{moffet1968minimum,pal2010nested,pal2011coprime}. Techniques such as spatial smoothing MUSIC (SS MUSIC)~\cite{liu2015remarks} or sparse recovery based DOA estimation techniques~\cite{vaidyanathan2010sparse} can be used on the co-array data computed from such sparse arrays.
	 
	In recent times, with the advances in MEMS technology, sophisticated microphones and transducers that are capable of measuring a vector quantity such as the acoustic particle velocity are becoming practically feasible~\cite{de2007microflown}. An acoustic vector sensor~(AVS) is one such device that is capable of measuring both the acoustic pressure and particle velocity at a given spatial location. It consists of an omni-directional microphone and two (or sometimes three) transducers, which are placed orthogonal to each other. These vector sensors are capable of measuring acoustic particle velocity in either $\mathbb{R}^{2}$ (or $\mathbb{R}^3$). Unlike a microphone (or a scalar sensor), a single AVS can measure the DOA of a far-field event~\cite{kitchens2010acoustic} and arrays of such AVSs have proven to have distinct advantages compared to conventional microphone arrays~\cite{nehorai1994acoustic}.
	This concept of vector sensor has also been extended to localize electromagnetic sources~\cite{nehorai1994vector}. 
	To estimate the DOAs with AVS arrays, we can adapt subspace algorithms that are developed for APS (i.e., scalar sensor) arrays~\cite{capon1969high, malioutov2005sparse}. Extensions to sparse AVS arrays and DOA estimation using co-array data models are also available~\cite{ han2014nested, rao2015sparse,krishnaprasadacoustic}.
	
	The DOA estimates obtained from these aforementioned standard algorithms for both the element-space and the covariance data models are highly sensitive to model errors~\cite{swindlehurst1992performance}, which mainly include relative gain and phase mismatches between different sensors within the array. We refer to gain and phase mismatches between different sensors as calibration errors. These calibration errors originate from the differences in the receiver/data acquisition electronics as well as the inherent uncertainties in the manufacturing of the transducers. The data acquisition electronics (e.g., oscillator and amplifier) of the sensors deviates from its nominal performance over a period of time and is also affected by the environmental conditions. As a result, they require periodic recalibration for which self calibration methods are useful.

	\subsection{State-of-the-art calibration methods}
	There is rich literature that explores techniques for DOA estimation in the presence of gain and phase uncertainties between the sensors~\cite{paulraj1985direction, astely1999spatial, ramamohan2018blind, weiss1990eigenstructure, liu2011eigenstructure, cao2013hadamard, wijnholds2009multisource, ling2016self}. Most of these approaches are self calibration techniques and blind in the sense that do not require calibrator sources with known direction and transmitted waveform. In~\cite{paulraj1985direction}, a blind self calibration procedure was presented for scalar sensor arrays arranged in a uniform linear array (ULA) configuration. In this approach the Toeplitz structure of the data covariance matrix was utilized by exploiting the redundancies in the uniform geometry to create an overdetermined system of equations and estimate the calibration parameters. In~\cite{wylie1994joint}, a calibration approach similar to~\cite{paulraj1985direction} but with a constraint on the phase errors was proposed and a constrained Cram\'{e}r-Rao bound for the self calibration problem was presented. Further in~\cite{li2006theoretical}, a simplified version of~\cite{paulraj1985direction} based on using only a few upper triangular entries of the covariance matrix was proposed. The extension of the approach in~\cite{paulraj1985direction} utilizing the redundancies (in the resulting Toeplitz block Toeplitz covariance matrix) of a uniform rectangular array (URA) was presented in~\cite{heidenreich2011gain}. The adaptation of eigenstructure methods to estimate sensor gain and phase uncertainties in~\cite{paulraj1985direction} to an AVS ULA was presented in~\cite{ramamohan2018blind}.
	
	In contrast to the above mentioned redundancy-based calibration approaches for regular geometries, in~\cite{weiss1990eigenstructure} an iterative algorithm to solve the nonconvex problem for simultaneously estimating both the DOAs and calibration parameters was proposed. This approach is applicable to arrays with arbitrary sensor geometries with a preference for non-linear geometries over linear arrays. In~\cite{wijnholds2009multisource}, a weighted alternating least squares~(WALS) approach was proposed for estimating the array parameters including the DOAs, source and noise powers, and calibration errors. However, these two existing approaches in~\cite{weiss1990eigenstructure}~and~\cite{wijnholds2009multisource} suffer from the choice of the initial estimate and the algorithm might only converge to a local minimum. Further, the conditions for a unique solution to estimate both the DOAs and calibration parameters with scalar sensor arrays was presented in~\cite{weiss1990eigenstructure}~and~\cite{astely1999spatial}.
	
	In the presence of large calibration errors, the one-step techniques presented in~\cite{cao2013hadamard}~and~\cite{liu2011eigenstructure} estimate the DOAs by considering modified measurements which are independent of phase errors. However, these techniques are applicable only to nonlinear arrays in the presence of two or more sources with limitations on the spatial separation between them.
	
	More recently, with the increased attention on sparse sensing, an algorithm for DOA estimation with sparse arrays (in particular, for nested arrays) under model/calibration errors was proposed in~\cite{han2015calibrating}. For gain error estimation they use the same approach as~\cite{paulraj1985direction}, while for the estimation of the phase errors a sub-optimal method based on the sparse total least squares (STLS) approach was proposed. 
	
	The self-calibration problem for linear models (not for DOA estimation) with a known sensing matrix was considered in~\cite{ling2016self}, wherein they convert the bilinear inverse problem to a linear problem. Then the obtained linear problem can be easily solved using a least squares approach. For the array processing problem, this means that the source directions are known and the inverse problem amounts to estimating the calibration parameters and source symbols. We draw inspiration from~\cite{ling2016self} to decouple the calibration parameters from the other unknowns, however, the model we deal with is not linear anymore as the source directions are not known. In essence, the main problem of interest in this work is self-calibration with non-linear measurement models.

	\subsection{Our contributions}
	
	Alternative to all the discussed approaches, in this work we propose unified algorithms/solvers for jointly estimating the source DOAs as well as the calibration parameters for both the scalar and vector sensor arrays using both the element-space and covariance domain data models. Also, the proposed algorithms are useful for arbitrary sensor geometries, such as uniform regular or sparse irregular linear arrays. We extend~\cite{ramamohan2019blind}, a precursor version of this paper, in several ways as summarized next.

	\begin{itemize}
	\item We develop novel self calibration algorithms for both the element-space and co-array data models, where the latter data model is useful when there are more sources than sparsely placed sensors. Addressing the calibration problem for sparse linear arrays with more sources than physical sensors has not received much attention expect for~\cite{han2015calibrating}, in which conditions required to obtain a unique solution were not presented. 
	\item For the aforementioned data models, we derive conditions to obtain a unique solution for estimating the DOAs and calibration parameters for vector sensor arrays and sparse APS arrays.
	\item One of the main results of this work is that AVS arrays require fewer reference sensors compared to their APS array counterparts for joint DOA estimation and calibration.
	\item To demonstrate the  developed methods, numerical experiments on synthetic data and experimental results on measurements carried out in an anechoic chamber are presented.
	\end{itemize}	
	
	\subsection{Notation and outline}
	Upper (lower) bold face letters are used for matrices (column vectors); $(\cdot)^{*}$ denotes conjugate, $(\cdot)^{T}$ denotes transpose and $(\cdot)^{H}$ denotes conjugate transpose; $\otimes$ denotes the Kronecker product, $\circ$ denotes the Khatri-Rao product and $\odot$ denotes the Schur-Hadamard (element-wise) product; $\mathbb{E}\{\cdot\}$ denotes the expectation operator; $\rm{tr}(\cdot)$ denotes the trace operator and $\mathbf{I}_{n}$ is the identity matrix of dimension $n$.

	The detailed outline for this paper is as follows. In Section~\ref{sec:datamodel}, we present the element-space and the covariance domain measurement model with calibration errors, and the problem statement of estimating both the calibration errors and the DOAs. In Section~\ref{sec:ambi_iden}, we discuss the ambiguity issues and its implications that arise while jointly estimating the calibration errors and the source DOAs. Further, for both the APS and AVS arrays, we present the identifiability conditions for uniquely and jointly estimating the calibration errors and the source DOAs. In Section~\ref{sec:calib_elementSpace}~and~Section~\ref{sec:calib_coarraySpace}, the proposed calibration algorithms based on the element-space and the covariance domain measurement data model are presented, respectively. The simulation results of the proposed calibration algorithms are discussed in Section~\ref{sec:NumericalExperiments}. Finally, in Section~\ref{sec:ExperimentalResults}, the details of the experiments performed in the anechoic chamber and its associated results are presented.

\section{Problem statement} \label{sec:datamodel}	
	Consider a linear array of $M$ sensors with $Q$ channels, where $Q = M$ for APS arrays and $Q = 3M$ for AVS arrays, for estimating the azimuth directions of $N$ narrow-band sources, denoted by $\boldsymbol{\theta} = [\theta_{1},  \theta_{2}, \ldots, \theta_{N}]^{T}$. The azimuth directions are with respect to the phase reference of the array. Each of the considered $Q$ channels has a different receiver gain and is not known. Let us denote the unknown gain of the $i$th channel as $g_i = \alpha_i e^{j\phi_i}$ with $\alpha_i$  and $\phi_i$ being the magnitude and phase, respectively. We collect these gains in the  diagonal matrix ${\rm diag}({\bf g})$ with ${\bf g} = [g_1,g_2,\ldots,g_Q]^T$. Let us also define the vectors ${\boldsymbol \alpha} = [\alpha_1,\alpha_2,\ldots, \alpha_Q]^T$ and ${\boldsymbol \phi} = [\phi_1,\phi_2,\ldots, \phi_Q]^T$. 

	Under the narrow-band assumption~\cite{opac-b1105852}, the element-space signal, ${\bf x}(t)$, can be modeled as~\cite{paulraj1985direction}
	\begin{equation}
	{\bf x}(t)= {\rm diag}({\bf g})\left[{\bf A}(\boldsymbol{\theta}) \, {\bf s}(t) + {\bf n}(t)\right] \in \mathbb{C}^{Q \times 1},
	\label{eq:data_model}
	\end{equation}
	where 
	\[
	{\bf A}(\boldsymbol{\theta}) =  \left[{\bf a}(\theta_1) \,\cdots \, {\bf a}(\theta_N)\right] \in \mathbb{C}^{Q \times N}
	\]
	is the array manifold matrix, the source signals of wavelength $\lambda$ are stacked in the vector ${\bf s}(t) \in \mathbb{C}^{N \times 1}$  and the receiver noise vector is given by $\mathbf{n}(t) \in \mathbb{C}^{Q \times 1}$. The location of the $m$th element of the array is denoted by $\delta_m$ with ${p}_m = \delta_m/\lambda$. Let us define $\mathbf{p} = [p_1,p_2,\ldots,p_M]^T$. Without loss of generality, we consider the first sensor with $p_1 = 0$ as the phase reference of the array.
	The spatial signature (or the array steering vector) for the $n$th source in the direction described by the vector $\mathbf{u}(\theta_{n}) = [\cos(\theta_{n}) \,\, \sin(\theta_{n})]^{T}$ with respect to the first sensor of the APS array with $M$ sensors is given by
	\begin{equation}
	{\bf a}_{\text{APS}}(\theta_{n}) =  \left[e^{\text{j}{2\pi}{p}_{1} \cos(\theta_n)}  \,\, \hdots \,\, e^{\text{j}{2\pi}p_M \cos(\theta_n)}\right]^{T} \in \mathbb{C}^{M \times 1},
	\label{eq:steeringAPS}
	\end{equation}
	whereas the related array steering vector of the AVS array is given by
	\begin{eqnarray}
	{\bf a}_{\text{AVS}}(\theta_{n}) &=&  \left[\begin{array}{l cr}  1 &  {\bf u}^T(\theta_{n}) \end{array}\right]^{T} \otimes {\bf a}_{\text{APS}}(\theta_n),\nonumber\\
	  &=& {\bf h}(\theta_{n}) \otimes 
	{\bf a}_{\text{APS}}(\theta_{n}) \in \mathbb{C}^{3M \times 1}. \label{eq:steeringAVS}
	\end{eqnarray}
	For the APS array, we have $Q=M$ with ${\bf a}(\theta_{n}) = {\bf a}_{\text{APS}}(\theta_{n})$ and for the AVS array we have $Q = 3M$ channels with ${\bf a}(\theta_{n}) = {\bf a}_{\text{AVS}}(\theta_{n})$. Usually the signal ${\bf x}(t)$ is uniformly sampled and $L$ snapshots are collected in the data matrix ${\bf X} = [{\bf x}(1), {\bf x}(2),\ldots,{\bf x}(L)] \in \mathbb{C}^{Q \times L}$ to obtain 
	\begin{equation}
	{\bf X} = {\rm diag}({\bf g})\left[{\bf A}(\boldsymbol{\theta}) {\bf S} + {\bf N}\right].
	\label{eq:data_model_matrix}
	\end{equation}
	Here, ${\bf S} = [{\bf s}(1), {\bf s}(2),\ldots,{\bf s}(L)] \in \mathbb{C}^{N \times L}$ and ${\bf N} = [{\bf n}(1), {\bf n}(2),\ldots,{\bf n}(L)] \in \mathbb{C}^{Q \times L}$. 
	The covariance matrix of the signal ${\bf x}(t)$ is ${\bf R}_{\rm x} = \mathbb{E}\{{\bf x}(t){\bf x}^H(t)\}  \in \mathbb{C}^{Q \times Q}$. We will assume that the source signals ${\bf s}(t)$ are uncorrelated and have a diagonal covariance matrix $\mathbb{E}\{{\bf s}(t){\bf s}^H(t)\} = {\rm diag}({\boldsymbol \sigma}_{\rm s})$, which is not known. Similarly, the noise vector has a diagonal covariance matrix $\mathbb{E}\{{\bf n}(t){\bf n}^H(t)\} = {\rm diag}({\boldsymbol \sigma}_{\rm n})$, which is assumed to be known or can be estimated. Then, the covariance domain model can be written as
	\begin{equation}
		{\bf R}_{\rm x} = {\rm diag}({\bf g})\left[{\bf A}(\boldsymbol{\theta}) {\rm diag}({\boldsymbol \sigma}_{\rm s}) {\bf A}^H(\boldsymbol{\theta})  + {\rm diag}({\boldsymbol \sigma}_{\rm n}) \right]{\rm diag}^H({\bf g}).	   
	\label{eq:covariance_matrix_yhat}
	\end{equation}
	Here, it is assumed that $\mathbf{s}(t)$ and $\mathbf{n}(t)$ are mutually uncorrelated. The covariance matrix ${\bf R}_{\rm x}$ has a Toeplitz structure for ULAs. When the array geometry is not uniform, it does not have any specific structure. It is also useful to express~\eqref{eq:covariance_matrix_yhat} in the vectorized form as:
	\begin{equation}
		{\bf r}_{\rm x} = {\rm diag}({\bf g}^{*} \otimes {\bf g})\left[{\bf A}_{\rm co}(\boldsymbol{\theta}){\boldsymbol \sigma}_{\rm s}   + {\boldsymbol \sigma}_{\rm n} \right],	   
	\label{eq:covariance_matrix_yhat_vecForm}
	\end{equation}
	where $\text{vec}(\mathbf{R}_{\rm x}) = \mathbf{r}_{\rm x}$ and ${\bf A}_{\rm co}(\boldsymbol{\theta}) = {\bf A}^{*}(\boldsymbol{\theta}) \circ {\bf A}(\boldsymbol{\theta})$ with the subscript ``${\rm co}$" indicating the co-array manifold. In practice, the data matrix ${\bf X}$ is used to compute the sample data covariance matrix $\widehat{\bf R}_{\rm x} =L^{-1} {\bf X}{\bf X}^H$. For the sake of convenience, further in this work, we will use ${\bf R}_{\rm x}$ instead of $\widehat{\bf R}_{\rm x}$ with the knowledge that only an estimate of the covariance matrix is available.
	
	Based on the co-array model in~\eqref{eq:covariance_matrix_yhat_vecForm}, the sensor elements can be smartly placed irregularly along the linear axis, such that ${\bf A}_{\rm co}$ has full column rank. Usually such configuration of linear arrays leads to sparse array design~\cite{pal2011coprime} allowing one to resolve as many as $\mathcal{O}(M^2)$ sources using $M$ sensors.
	
	The main goal of this paper is to jointly estimate the $Q$ complex (i.e., $2Q$ real) receiver gains ${\bf g}$ and $N$ directions ${\boldsymbol \theta}$ given ${\bf X}$ or ${\bf r}_{\rm x}$. To do so uniquely, as will be discussed in Section~\ref{sec:ambi_iden}, we will require a few reference sensors with known complex receiver gains in the array.

	\section{Ambiguity and identifiability}
	\label{sec:ambi_iden}

	Before presenting the calibration algorithms, in this section, we discuss identifiability conditions under which a unique solution for both the calibration parameters and the source DOAs exists. We will do this for calibration techniques based on both the element-space data model~\eqref{eq:data_model_matrix} and co-array data model~\eqref{eq:covariance_matrix_yhat_vecForm}, and for both APS and AVS arrays. Specifically, we focus on linear arrays. It should be immediately clear that, as both ${\rm diag}({\bf g}){\bf A}({\boldsymbol \theta})$ and ${\bf S}$ (or ${\boldsymbol \sigma}_{\rm s}$) are not known a priori, they cannot be computed uniquely as there will be a complex (or real) scaling ambiguity. Therefore, to fix the scaling ambiguity we perform calibration with respect to {\it sensor 1} at location $p_1 = 0$, i.e., we use $g_1 = 1$ for the element-space data model and $|g_1| = \alpha_{1} = 1$ for the co-array data model. 
	
	After establishing the fact that the elements of $\bf{g}$ can only be estimated relative to the reference sensor, the next important question that needs to be addressed is under what conditions can we uniquely estimate ${\bf g}$ and ${\boldsymbol \theta}$ given the measurement data. However due to the bi-linear nature of the estimation problem, it is not straightforward to derive the identifiability conditions based on the element-space data model~\eqref{eq:data_model_matrix} or co-array data model~\eqref{eq:AVS_Msensor_Nsource_sparse_model_opt_problem_convex_trace_constraint}. Therefore, we adapt the approach discussed in~\cite{weiss1990eigenstructure}, to derive the sufficient conditions for uniquely estimating $\bf{g}$ and $\boldsymbol{\theta}$ based on the assumption that ${\rm diag}({\bf g}) \mathbf{A} ({\boldsymbol \theta})$ (${\rm diag}({\bf g}^{*} \otimes {\bf g}){\bf A}_{\rm co}(\boldsymbol{\theta})$) is given, with the knowledge that only the column span of it is available from the measurement data.
	
	\subsubsection{The element-space data model}	
	From the element-space data model~\eqref{eq:data_model_matrix}, we have $2QL$ nonlinear equations in $N$ unknown DOAs, $2(Q-1)$ unknown calibration parameters, and $2NL$ unknown source signals. Hence, for well posedness of the calibration problem, we require
	\[
	2QL \geq N + 2Q-2 +2NL \quad \Rightarrow \quad  \frac{N + 2(Q -1)}{2(Q-N)} \leq L,
	\]
	which is meaningful only for $Q > N$. For deriving the sufficient conditions, let us define the phase of ${\rm diag}({\bf g}) \mathbf{A} ({\boldsymbol \theta})$ as
	\begin{equation}
	\rho_{q}(n) = \frac{1}{2\pi}{\rm angle}\left(g_q \left[{\bf A}({\boldsymbol \theta})\right]_{qn}\right) =  p_{q} \cos(\theta_{n}) + \phi_{q},
	\label{eq:phaseGainTerm_modAmatrix}
	\end{equation}
	for $q = 1,\hdots,Q$ and $ n = 1,\hdots, N.\,$ Introducing $\boldsymbol{\rho}_{n} = \left[\rho_{1}(n), \hdots, \rho_{Q}(n)\right]^{T}$ and defining $\mathbf{p}_{\rm ext} :=  {\bf p}$ for the APS array and $\mathbf{p}_{\rm ext} :=  {\bf 1}_3 \otimes {\bf p}$ for the AVS array, we can write the above equation compactly as 
	\begin{equation}
	\label{eq:phaseRelationsElementSpaceModel}
	\boldsymbol{\rho}_{n} 
	= \mathbf{p}_{\rm ext}\, \cos(\theta_{n})  + {\boldsymbol{\phi}} 
	= \begin{bmatrix}\mathbf{p}_{\rm ext} & \mathbf{I}_{Q}\end{bmatrix}\, \begin{bmatrix} \cos(\theta_{n}) \\ {\boldsymbol{\phi}}\end{bmatrix}
	\end{equation}
	for $n = 1, 2, \ldots, N$. This is an under-determined system of $Q$ equations, which has rank $Q-1$ (with $\phi_1 = 0$), and $Q$ unknowns. It is possible to solve~\eqref{eq:phaseRelationsElementSpaceModel}, if another sensor/channel's phase error is known in the array (say w.l.o.g. $\phi_2 = 0$). However, when $N \geq 2$, we can eliminate $\boldsymbol{\phi}$ by considering
	\[
	\boldsymbol{\rho}_{n} - \boldsymbol{\rho}_{1} =\mathbf{p}_{\rm ext} \left[\cos(\theta_{n}) - \cos(\theta_{1})\right],
	\]
	to obtain $N-1$ linearly independent equations in $N$ unknown DOAs of the form
	\begin{equation}
	\label{eq:VectorPhaseTermDiff_modAmatrix}
	\mathbf{p}_{\rm ext}^{\dagger} \,\, \left(\boldsymbol{\rho}_{n} - \boldsymbol{\rho}_{1}\right) = \cos(\theta_{n}) - \cos(\theta_{1}); \quad n = 2, \cdots,N.
	\end{equation} 	

	The above system in~\eqref{eq:VectorPhaseTermDiff_modAmatrix} is still underdetermined. Nonetheless, if one of the DOAs is known (say, $\theta_1$ is known w.l.o.g.,) then we can identify the remaining DOAs. This result for a scalar sensor array ($Q = M$) was presented in~\cite{weiss1990eigenstructure}. 

	Interestingly for an AVS array ($Q = 3M$), the need of knowing the direction of one calibrator source $\theta_1$ can be relaxed as the direction information is available in the magnitude of the element-space data model. Considering only the magnitude of  ${\rm diag}({\bf g}) \mathbf{A} ({\boldsymbol \theta})$, we have 
	\begin{equation}
	\label{eq:gainTerm_modAmatrix}
	\nu_{q}(n) = \left|g_q\left[\mathbf{A} ({\boldsymbol \theta})\right]_{qn}\right| = \alpha_{q} \left|h_{q}(\theta_n)\right|
	\end{equation}
	for $q = 1,\ldots, 3M$ and $n = 1,\ldots, N$. Here,
	\begin{equation}
	\label{eq:hAVS}
	h_{q}(\theta_n) = \begin{cases}1, & 1\leq q\leq M.\\ \cos(\theta_{n}), & M+1\leq q\leq 2M. \\ \sin(\theta_{n}), & 2M+1\leq q\leq 3M.\end{cases}
	\end{equation}
	Let us consider the equations related to $q = M+1$, which are given by
	\[
	\nu_{M+1}(n) = \alpha_{q+1} \cos(\theta_n).
	\]
	As we assume $N \geq 2$, we can eliminate the unknown $\alpha_{q+1}$ to obtain 
	\[
	\cos(\theta_1) = \frac{\nu_{M+1}(1)}{\nu_{M+1}(n)} \cos(\theta_n) . 
	\]
	Thus we can compute $\theta_1$ as
	\[
	\theta_{1} = {\arccos}\left(\frac{\nu_{M+1}(n)}{\nu_{M+1}(n)}   \cos(\theta_{n})\right).
	\]
	This value of $\theta_{1}$ can be used in~\eqref{eq:VectorPhaseTermDiff_modAmatrix}, which eliminates the need for knowing one of the DOAs for uniquely identifying all the $N$ DOAs for the AVS linear array. The array manifold matrix ${\bf A}({\boldsymbol \theta})$ is known once all the $N$ DOAs are computed. Then using~\eqref{eq:phaseGainTerm_modAmatrix} and~\eqref{eq:gainTerm_modAmatrix}, respectively, the phase and gain errors can be computed.


	Now to check if the derived sufficient condition for the APS linear array is also necessary, we need to show that the solution of $\bf{g}$ and $\boldsymbol{\theta}$ is not unique if we do not consider the calibrator source. To do so, we provide a counter example by considering an $M$-element APS ULA, and $N$ far-field sources. For such configuration, due to the Vandermonde structure of ${\bf A}({\boldsymbol \theta})$, we can have ${\rm diag}({\bf g}){\bf A}(\boldsymbol{\theta}) = {\rm diag}({\bf g} \odot {\bf a}(\theta_0))({\bf A}(\boldsymbol{\theta}) \odot {\bf a}^{*}(\theta_0))$ = ${\rm diag}(\tilde{{\bf g}}){\bf A}(\tilde{\boldsymbol{\theta}})$, where ${\bf g} \neq \tilde{\bf g}$ and ${\boldsymbol \theta}  \neq  \tilde{\boldsymbol \theta}$ indicating the non-uniqueness of the solution.
	\\
	
	{\centering 
		\fbox{\begin{minipage}{1\textwidth} \textbf{\textit{Based on the element-space formulation for a linear APS array, irrespective of the array geometry, given $Q > N$, $N\geq2$ and ${\rm diag}(\bf{g}){\bf A}({\boldsymbol \theta})$, the requirement of a calibrator source is a sufficient and necessary condition for a unique solution of ${\bf g}$ and $\boldsymbol{\theta}$ to exist. On the other hand there is no requirement of calibrator source for linear AVS array.}}\end{minipage}}}
	\vspace{0.2cm}

	\subsubsection{The co-array data model} \label{ref:coarrayIden}
	In the co-array data model in~\eqref{eq:covariance_matrix_yhat_vecForm}, we have $2QN - N^2 +1$ nonlinear equations\footnote{The covariance matrix ${\bf R}_{\rm x}$ is completely characterized by $N+1$ real eigenvalues and $2QN - N^2 -N$ real parameters related to the orthonormal eigenvectors associated to the sources.} in $N$ unknown DOAs, $2(Q-1)$ unknown calibration parameters, and $N$ unknown source powers. Hence, for well posednes,s we require
	\[
	2QN - N^2 +1 \geq 2N + 2Q-2  \quad \Rightarrow \quad   Q \geq  \frac{N^2 + 2N - 3}{2(N-1)}.
	\]
	Next, we discuss the sufficient conditions to estimate ${\bf g}$ and ${\boldsymbol \theta}$, given ${\rm diag}({\bf g}^{*} \otimes {\bf g}){\bf A}_{\rm co}(\boldsymbol{\theta})$. 
	To do so, consider the phase of ${\rm diag}({\bf g}^{*} \otimes {\bf g}){\bf A}_{\rm co}(\boldsymbol{\theta})$ that is given by 
	\begin{eqnarray}
	\rho_{pq}(n) &=& \frac{1}{2\pi}\text{angle}\left(g_p^* g_q \left(\left[{\bf A}^{*}(\boldsymbol{\theta})\right]_{pn}\circ\left[{\bf A}(\boldsymbol{\theta})\right]_{qn}\right)\right)\nonumber\\ 
	&=&  \left(p_{p} - p_{q}\right) \cos(\theta_{n})
	- \left(\phi_{p} - \phi_{q}\right),
	\label{eq:phaseGainTerm_modAmatrix_CovarModel}
	\end{eqnarray}
	for $p,\,q = 1,\cdots, Q$ with $p\neq q$ and $n = 1,\cdots, N.$
 
	If $N = 1$, we require two sensors/channels with known phase errors. Suppose w.l.o.g., that $\phi_1 = \phi_2 =0$, then we can compute the DOA as
	\begin{equation}
	\label{eq:theta1cov}
		\theta_1 = \arccos\left(\frac{\rho_{12}(1)}{p_1 -p_2}\right),
	\end{equation}
	with no specific requirements for $p_1 \neq 0$ or $p_2 \neq 0$. Defining $\boldsymbol{\rho}_{n} = [\rho_{11}(n), \rho_{12}(n), \ldots, \rho_{QQ}(n)]^T$, we can compactly write~\eqref{eq:phaseGainTerm_modAmatrix_CovarModel} as
	\begin{equation}
	\label{eq:phaseRelationsVecForm}
		\boldsymbol{\rho}_{n} = {\bf D}{\mathbf{p}_{\rm ext}}\, \cos(\theta_{n}) - {\bf D}{\boldsymbol{\phi}} = \begin{bmatrix}{\bf D}{\mathbf{p}_{\rm ext}} &  -{\bf D}\end{bmatrix}\begin{bmatrix}\cos(\theta_{n}) \\ {\boldsymbol{\phi}}\end{bmatrix}, 
	\end{equation}
	where ${\bf D} \in  \mathbb{R}^{Q^{2}-Q}$ is the difference matrix that we use to compute the pairwise differences in~\eqref{eq:phaseGainTerm_modAmatrix_CovarModel}. If $N \geq 2$, irrespective of the array geometry, the phase errors $\boldsymbol{\phi}$ can be eliminated by considering
	\begin{equation}
	\boldsymbol{\rho}_{n} - \boldsymbol{\rho}_{1} = \mathbf{D} {\mathbf{p}_{\rm ext}}\, \left[\cos(\theta_{n}) - \cos(\theta_{1})\right]; \quad \forall{n} = 2, \hdots, N,\nonumber
	\end{equation}
	which can be equivalently expressed as
	\begin{equation}
	\theta_{n}  = \arccos \left( \left(\mathbf{D}{\mathbf{p}_{\rm ext}}\right)^{\dagger} [ \boldsymbol{\rho}_{n} - \boldsymbol{\rho}_{1}] + \cos(\theta_{1})\right).
	\label{eq:phaseRelBasedCovar}
	\end{equation}
	This is similar to the element-space version as seen in~\eqref{eq:VectorPhaseTermDiff_modAmatrix} and it is underdetermined. Nonetheless for APS linear arrays, similar to the element-space model, if one of the source DOAs is known (say, $\theta_1$ is known w.l.o.g.) then we can identify the remaining DOAs.
	
	For AVS array, similar to the element-space model, the magnitude of ${\rm diag}({\bf g}^{*} \otimes {\bf g}){\bf A}_{\rm{co}}(\boldsymbol{\theta})$ also contains the direction information. Specifically,
	\begin{eqnarray}
		\nu_{pq}(n) &=& \left| g_p g_q\left(\left[{\bf A}^{*}(\boldsymbol{\theta})\right]_{pn}\circ\left[{\bf A}(\boldsymbol{\theta})\right]_{qn}\right)\right|,\nonumber\\ 
		&=& \psi_{p}  \psi_{q}  h_{p}(\theta_n) h_q(\theta_n)
	\label{eq:amplitudeRelationsCovarAVS}
	\end{eqnarray}
	for $p,q = 1,\cdots, 3M$, where we recall that $h_{p}(\theta_n)$ is as in \eqref{eq:hAVS} and $\theta_n \in [0,\pi]$ for $n = 1,\hdots, N$.
	Consider w.l.o.g., the equation related to $p = M+1$ and $q = M+2$, i.e.,
	\[
	\nu_{M+1\,M+2}(n) = \psi_{M+1}  \psi_{M+2} \cos^2(\theta_n).
	\]
	When $N \geq 2$, we can eliminate the unknown gain errors $\psi_{M+1}$ and  $\psi_{M+2}$ above as	
	\[
	\cos(\theta_1) = \left[ \frac{\nu_{M+1\,M+2}(1)}{\nu_{M+1\,M+2}(n)} \cos^2(\theta_n) \right]^{1/2},
	\]
	which can now be used in \eqref{eq:phaseRelBasedCovar} to compute the DOAs. Once the DOAs are computed for either the APS or AVS array, the phase errors can be computed from~\eqref{eq:phaseRelationsVecForm}, with respect to one of the reference sensors/channels in the array as the rank of ${\bf D}$ is always $Q-1$. The gain errors can be computed from the amplitude relations in~\eqref{eq:amplitudeRelationsCovarAVS}.
	\\
	
	{\centering 
		\fbox{\begin{minipage}{1\textwidth} \textbf{\textit{
	It can be concluded that irrespective of the array geometry of the linear array with the co-array data model, it is sufficient to have one phase reference sensor and one~(no) calibrator source for a linear APS~(AVS) array, for uniquely estimating ${\bf g}$ and ${\boldsymbol \theta}$ when $N \geq 2$.}}\end{minipage}}}
	\vspace{0.2cm}
	
	Similar to the element-space approach, we see that one calibrator source is required for an APS linear array for uniquely estimating $\bf{g}$ and $\boldsymbol{\theta}$. However, unlike~\eqref{eq:phaseRelationsElementSpaceModel}, which is an under-determined system, it can be observed that~\eqref{eq:phaseRelationsVecForm} is a tall system with $(Q^2 - Q)$ equations and $(Q+1)$ unknowns. APS linear arrays with a particular structure in the array geometry, such as specific sparse arrays or uniform linear arrays (ULAs) result in redundant relations that are part of~\eqref{eq:phaseRelationsVecForm}. Those redundancies in the structured APS linear array allow for estimating $\bf{g}$ and subsequently $\boldsymbol{\theta}$ without knowledge of a known calibrator source leading to another set of sufficient conditions. This is discussed in the following part.
	
	From the co-array perspective of scalar sensor arrays, the distinct elements of $\mathbf{D}\mathbf{p}_{\rm{ext}}$, as seen in~\eqref{eq:phaseRelationsVecForm}, behave like virtual sensor locations given by the difference set~$\{p_i - p_j, 1\leq i,j \leq M\}$. Those virtual sensor locations increase the degrees-of-freedom (DOF) of the array allowing for estimating more sources than physical sensors, if they are placed strategically. In order to look at the self-calibration problem for such array configurations, let us reuse some definitions from~\cite{pal2010nested}.
	
	\begin{mydef}
		\textbf{(Difference co-array)}
		For an $M$-element sensor array, with $p_i$ denoting the position of the $i$th sensor, define the set
		$$\mathcal{D} = \{p_i - p_j\}, \,\, \forall i,j = 1,2,\hdots,M,$$
		which allows for a repetition of its elements. We also define the set $\mathcal{D}_{\mathcal U}$, which consists of the distinct elements of the set $\mathcal{D}$. Then, the \textit{difference co-array} of the given array is defined as the array which has sensors located at positions given by the set $\mathcal{D}_{\mathcal U}$.
	\end{mydef}
		
	\begin{mydef}
		\textbf{(Weight function)}
		An integer valued weight function $w:\mathcal{D}_{\mathcal U} \rightarrow \mathbb{N}^{+}$ is defined as 
		$$w(p) =\text{no. of occurances of}\,\,p\,\,\text{in}\, \mathcal{D},\,p \in \mathcal{D}_{\mathcal U},$$
		where $\mathbb{N}^{+}$ is the set of positive integers. The weight function $w(p)$ denotes the number of times $p$ occurs in $\mathcal{D}$.
	\end{mydef}	
		
	The cardinality of the set $\mathcal{D}_{\mathcal U}$ for a given array gives the degrees of freedom (DOF) that can be obtained from the difference co-array associated with that array. The motivation of sparse array design, such as the minimum redundancy array (MRA), sparse ruler array or nested array, is to maximize the number of DOF of the co-array, which in other words means the value of the weight function $w(p),\, \forall p \in \mathcal{D}_{\mathcal U} \setminus \{0\}$ has to be minimized. However, from the self calibration perspective a value of the weight function $w(p),\, \forall p \in \mathcal{D}_{\mathcal U} \setminus \{0\}$ greater than 1 is beneficial as this results in redundancies in~\eqref{eq:phaseRelationsVecForm}. By exploiting redundancies in those relations for a $n^{\rm th}$ source and each $p$, the directional terms can be eliminated resulting in an equation with only the phase terms, i.e., 
	\begin{eqnarray}
		\label{eq:diffPhaseRelations}
			\rho_{pq}(n) - \rho_{kl}(n) &=& \rho_{pqkl}(n) =   \phi_{p} - \phi_{q} - \phi_{k} + \phi_{l},
	\end{eqnarray}		
	where $p_p-p_q = p_k-p_l$ for $p,q,k,l = 1,\hdots,M$ and $n= 1,\hdots, N$. Such relations for all $p,q,k,l$ can be expressed as a system of equations, i.e.,
	\begin{eqnarray}
	\label{eq:sysOfEqns_PhaseRelations}
	\begin{bmatrix}\hdots & \rho_{pqkl}(n) & \hdots \end{bmatrix}^{T} =  {\bf G}\, \begin{bmatrix} \phi_1 & \hdots & \phi_M  \end{bmatrix}^{T},
	\end{eqnarray}
	where ${\bf G}$ is a deterministic matrix, which depends on the chosen array geometry and the phase errors can be estimated by inverting it. We now look into the rank of the ${\bf G}$ matrix for different structured linear arrays and summarize how the phase errors can be estimated for each of those scenarios,
	\begin{itemize}
		\item The maximum amount of redundancies can be found in a uniform linear array~(ULA), where for $M$ elements, $w(\pm{d}) = M-d,$ for $d = 0,1,\hdots,M-1$. The rank of ${\bf G}$ is then always $M-2$, indicating that the phase errors can be estimated with respect to an arbitrary reference and within an arbitrary progressive phase factor~\cite{paulraj1985direction}. 
		If two reference sensors with known phase errors are present in the ULA, then it is possible to calibrate the remaining sensors in the array with respect to those references or in other words it is possible to estimate all the elements in $\boldsymbol{\phi}$. A similar extension for an AVS ULA was presented in~\cite{ramamohan2018blind,weiss2020asymptotically}, where the rank of ${\bf G}$ is always $3M-2$ for an $M$-element array.
		\item
		To design an $M$-element sparse array, taking self calibration into consideration, there is a trade-off between DOF and redundancies. The maximum rank of ${\bf G}$ for an $M$-element APS array is upper bounded by $M-2$. The rank of ${\bf G}$ for a structured sparse arrays including 
			\begin{itemize}
				\item the nested array~\cite{pal2010nested} and super nested array~\cite{liu2016super} is always $M-3$, 
				\item the co-prime arrays~\cite{pal2011coprime}, which enjoy more redundancies, it is $M-2$.
			\end{itemize}
		\item If there is a provision to introduce additional sensors within a sparse array to allow for sufficient redundancies, then fewer reference sensors with known phase errors are needed. For example, for an MRA~\cite{moffet1968minimum}, in most of the cases, it is seen that if we can introduce two phase reference sensors in the array, then it is possible to calibrate all the sensors in the array with respect to them, i.e. for example,
		\begin{itemize}
			\item With $M = 5$ with ${\bf p} = [0, 1, 4, 7, 9]^T$, the rank of ${\bf G}$ is 1. If we introduce two phase reference sensors with ${\bf p} = [0, 1, \textbf{2, 3}, 4, 7, 9]^T$, then the rank of ${\bf G}$ is 5.
		\end{itemize}
	\end{itemize}

	On the other hand, the gain errors can be estimated with respect to the chosen reference sensor~(i.e., to fix the scaling ambiguity) with known gain error by considering the amplitude relations of ${\rm diag}({\bf g}^{*} \otimes {\bf g}){\bf A}_{\rm co}(\boldsymbol{\theta})$. Unlike the estimation of the phase errors, estimating the gain errors using redundancies is applicable to all the linear arrays irrespective of its geometry~\cite{paulraj1985direction}.

	\vspace{0.4cm}
	{\centering 
	\fbox{\begin{minipage}{1\textwidth} \textbf{\textit{Using redundancy-based calibration techniques for an APS array based on the co-array data model, it can be concluded that for a ULA we need two phase reference sensors in the array while for sparse arrays we need at least two or more phase reference sensors in the array for uniquely estimating the calibration errors and source DOAs.}}\end{minipage}}}
	\vspace{0.2cm}
	
	\begin{myrem} To derive the sufficient conditions based on the redundancy-based calibration technique, we choose to have reference sensors with known phase errors in the array to improve the rank of the ${\bf G}$ matrix such that the phase errors can be estimated. However, we can also have other a priori conditions on the phase errors, such as $\sum_{q = 1}^{Q} \phi_q = 0$, that improve the rank of the ${\bf G}$, leading to another set of sufficient conditions to estimate the phase errors and subsequently the source DOAs uniquely.  \end{myrem}
		
	\begin{myrem} The identifiability conditions for non-linear AVS arrays can be derived along similar lines of non-linear APS arrays as in~\cite{weiss1990eigenstructure}. It can be shown that for both APS and AVS arrays with $N \geq 2$, it is sufficient to have one reference sensor with a known gain and phase error for uniquely estimating both the calibration parameters and the source DOAs. In particular for APS non-linear arrays, the need for a reference source can be eliminated for the purpose of calibration due to the presence of extra degrees-of-freedom in its spatial frequencies.\end{myrem}

	\section{Self calibration with the element-space model} \label{sec:calib_elementSpace}
	In this section, we focus on estimating the complex-valued receiver gains and the source DOAs, when only a few snapshots or a single snapshot is available. In such cases, the sample data covariance matrix will be a very poor estimate of ${\bf R}_{\rm x}$ and hence we focus on the element-space data model.  We begin with a simple scenario, wherein the source directions and the related signals are known, e.g., these could be calibrator sources. Later, we consider the joint estimation problem of interest without any calibrator source. The algorithms provided in this section, do not make any assumptions on the array geometry or on the structure of the covariance matrix~${\bf R}_{\rm x}$.
		
	Defining the diagonal calibration matrix ${\rm diag}({\bf c}) = {\rm diag}^{-1}({\bf g})$, we can express the ``calibrated" output signal ${\bf y}(t)$ as
	\begin{equation}
	\label{eq:linear_model}
		{\bf y}(t) = {\rm diag}({\bf c}){\bf x}(t) = {\rm diag}({\bf x}(t)){\bf c} = {\bf A}(\boldsymbol{\theta}) \, {\bf s}(t) + {\bf n}(t).
	\end{equation}
	Assuming that the true directions are from a uniform grid of $D \gg N$ points i.e., assuming that $\theta_n \in \left\{0, \frac{\pi}{D}\cdots, \frac{\pi(D-1)}{D}\right\}$, for $n = 1,2,\ldots, N$, we can approximate \eqref{eq:linear_model} as
	\begin{equation}
	\label{eq:linmod_approx}
	{\bf y}(t) = {\rm diag}({\bf c}){\bf x}(t) = {\rm diag}({\bf x}(t)){\bf c} = {\bf A}_{\mathbb{D}} \, {\bf z}(t) + {\bf n}(t),
	\end{equation}
	where ${\bf A}_{\mathbb{D}}$ is a $Q \times D$ dictionary matrix that consists of column vectors of the form
	${\bf a}(\bar{\theta}_d)$, with $\bar{\theta}_d$ being the $d$th point of the uniform grid of directions, i.e., $ \bar{\theta}_d = \frac{\pi d}{D}$, $d=0,1,\ldots, D-1$, and  ${\bf z}(t)$ is a length-$Q$ vector containing the source signal related to the corresponding discretized directions. We emphasize here that finding the columns of ${\bf A}_{\mathbb{D}}$ that correspond to non-zero elements of ${\bf z}(t)$ amounts to finding the DOAs.

	\subsection{Known calibration sources} \label{subsec:single_snapshot_scenario}
	Assuming we have a single snapshot, i.e., $L=1$, the data model \eqref{eq:linmod_approx} simplifies to
	\[
	{\bf y}(1)  = {\rm diag}({\bf x}(1)){\bf c} = {\bf A}_{\mathbb{D}} \, {\bf z}(1) + {\bf n}(1).
	\]
	The above system of equations may be written as a linear model in the unknowns ${\bf c}$ and ${\bf z}(1)$ as
	\begin{eqnarray}
	\begin{bmatrix}{\rm diag}({\bf x}(1))  &  -{\bf A}_{\mathbb{D}} \end{bmatrix}
	\begin{bmatrix}{\bf c} \\  {\bf z}(1) \end{bmatrix}   &=& {\bf n}(1). 
	\label{eq:linmod_T1}
	\end{eqnarray}
	This system is clearly underdetermined, and also from \eqref{eq:linear_model}, it is evident that the number of available equations, $Q$, is much less than the number of unknowns, $Q+D$. 
	
	When all the source directions and signals are known, i.e., when ${\bf z}(1)$ is known, then we use that knowledge to calibrate the array using simple least squares as we now have $Q$ equations in $Q$ unknowns. The calibration estimates are then simply given by 
	\[
	[\widehat{\bf c}]_q = \frac{\left[{\bf A}_{\mathbb{D}}{\bf z}(1)\right]_q}{\left[{\bf x}(1)\right]_q}, \,q=1,\ldots, Q.  
	\]
	This approach is a \textit{non-blind approach} where the prior information about the source locations and signals is used for array calibration without any other reference sensors. 	

	\subsection{Without calibration sources}
	
	When no calibration sources are available, we have seen that the system in \eqref{eq:linmod_T1} is under-determined. Nonetheless, leveraging the fact that the calibration parameters remain unchanged during an observation window where we collect $L$ snapshots, we can obtain more equations. To see this, we further develop \eqref{eq:linmod_T1} for multiple snapshots as
	\begin{equation}
	\underbrace{{\left[\begin{array}{c|cccc}{\rm diag}({\bf x}(1)) & -{\bf A}_{\mathbb{D}} &   &   &   \\\vdots &   &   & \ddots &   \\{\rm diag}({\bf x}(L)) &   &   &   & -{\bf A}_{\mathbb{D}}\end{array}\right]}}_{\bf G} \underbrace{{\left[\begin{array}{c}{\bf c} \\\hline {\bf z}\end{array}\right]}}_{\boldsymbol \gamma} = \underbrace{\left[\begin{array}{c}{\bf n}(1)\\  \vdots \\ {\bf n}(L)\end{array}\right]}_{\bf n},
	\label{eq:mmv}
	\end{equation}
	where ${\bf z} = \vec({\bf Z}) \in \mathbb{C}^{DL}$ with ${\bf Z} = [{\bf z}(1), {\bf z}(2), \cdots, {\bf z}(L)] = [{\bf z}_1, {\bf z}_2, \ldots, {\bf z}_D]^T.$ Here, ${\bf z}(l) \in \mathbb{C}^{D}$ and ${\bf z}_d \in \mathbb{C}^{L}$.  
	
	Although at the outset, it seems as if there are $Q+DL$ unknowns in \eqref{eq:mmv}, the vector ${\bf z}$ is structured. Specifically, the vectors ${\bf z}(l),\, l=1,\ldots,L$ are sparse, and more importantly, they have the \emph{same} sparsity pattern with the indices of the nonzero pattern indicating the source directions. The prior knowledge of having sparsity along the spatial domain can be incorporated by initially considering the $l_2$ norm of all the time samples corresponding to a particular spatial index of ${\bf Z}$, i.e., by defining $z_{d}^{(\ell_2)} = \| {\bf z}_d\|_2$ for $d = 1,2,\hdots,D$, and then by using the sparsity promoting $l_1$ norm penalty on the vector ${\bf z}^{(\ell_2)} = \left[z_{1}^{(\ell_2)},\,z_{2}^{(\ell_2)},\, \hdots, z_{D}^{(\ell_2)} \right]^{T}$ as~$f({\bf z}) = \| {\bf z}^{(l_{2})}\|_{\ell_1} =~\sum\limits_{d=1}^D z_{d}^{(\ell_2)}$.
	
	The optimization problem to jointly estimate the calibration parameters and DOAs with a sparsity constraint along the spatial domain of the matrix ${\bf Z}$ can then be expressed as:
	\begin{equation}
	\begin{aligned}
	&\underset{{\bf c}, \,{\bf z}}{\rm min} \quad   \|\mathbf{G}\boldsymbol {\gamma}\|_{2}^{2} + \lambda f({\bf z}) \quad \text{s.t.} \quad ({\bf c}, \,{\bf z}) \, \in \, \mathcal{C}
	\end{aligned}
	\label{eq:DetModel}
	\end{equation}
	where $\boldsymbol{\gamma} = [{\bf c}^T \,\, {\bf z}^T]^T$, $\lambda$ is the regularization parameter that allows for a trade off between the goodness of fit of the solution to the given data and the sparsity prior on ${\bf z}$. For choosing the right regularization parameter we can use many of the existing techniques from the compressive sensing techniques~\cite{malioutov2005sparse,hansen1992analysis}. The constraint set for APS arrays is $\mathcal{C} := \{({\bf c}, \,{\bf z}) \,\, | \,\,  c_1 = 1, {\bf z}_1 = {\bf 1}\}$ while for AVS arrays it is $\mathcal{C} := \{({\bf c}, \,{\bf z}) \,\, | \,\,  c_1 = 1\}$. Recall that for APS arrays, we need one reference sensor and we need to know one of the DOAs to avoid ambiguities. This is done by setting $c_1 = 1$ and  ${\bf z}_1 = {\bf 1}$, which is equivalent to having a calibrator source at $\bar{\theta}_1$ (w.l.o.g.). Since for AVS arrays, we do not need any calibrator source, we only need a reference sensor in that case. The optimization problem \eqref{eq:DetModel} is a convex optimization problem, which can be solved using any off-the-shelf solver. For large $L$, if the number of sources can be estimated, the complexity of the formulation in~\eqref{eq:DetModel} can be reduced by using the $\ell_1$-SVD technique~\cite{malioutov2005sparse} on the measurement data matrix ${\bf X}$. In other words, for $N$ sources, by just considering the basis of the signal subspace consisting of dimension $N \ll L$, the dimensionality of the measurement data matrix ${\bf X}$ can be reduced. Based on this reduced measurement data matrix the formulation in~\eqref{eq:DetModel} can be easily adapted.

	\section{Self calibration with the co-array data model} \label{sec:calib_coarraySpace}
	In this approach both the calibration errors (gain and phase errors) and the source DOAs will be estimated jointly based on the covariance matrix of the measurement data. The vectorized version of the covariance matrix in~\eqref{eq:covariance_matrix_yhat_vecForm} can also be expressed as
	\begin{equation}
		{\rm diag}({\bf c}^{*} \otimes {\bf c}){\bf r}_{\rm x} = {\bf A}_{\rm co}(\boldsymbol{\theta}){\boldsymbol \sigma}_{\rm s}   + {\boldsymbol \sigma}_{\rm n},	   
	\label{eq:coMatVecFormModified}
	\end{equation}
	where ${\rm diag}({\bf c}^{*} \otimes {\bf c}) = {\rm diag}^{-1}({\bf g}^{*} \otimes {\bf g})$.	Similar to~\eqref{eq:linmod_approx}, the directions can be assumed to be derived from a uniform grid of $D \gg N$ points. Then~\eqref{eq:coMatVecFormModified} can be approximated as
	\begin{equation}
	{\rm diag}({\bf c}^{*} \otimes {\bf c}){\bf r}_{\rm x} = {\rm diag}({\bf r}_{\rm x}) ({\bf c}^{*} \otimes {\bf c}) = {\bf A}_{{\rm co}\mathbb{D}}{\boldsymbol \sigma}_{\rm z}   + {\boldsymbol \sigma}_{\rm n},	   
	\label{eq:coMatVecFormModified_approx}
	\end{equation}
	where ${\bf A}_{{\rm co}\mathbb{D}}$ is a ${Q^2 \times D}$ dictionary matrix that consists of column vectors of the form ${\bf a}^{*}(\bar{\theta_{d}}) \otimes {\bf a}(\bar{\theta_{d}})$, with $\bar{\theta_{d}}$ as defined before. It can be easily observed that $({\bf c}^{*} \otimes {\bf c}) = \text{vec}({\bf C})$, with ${\bf C} = {\bf c}{\bf c}^{H}$, and hence~\eqref{eq:coMatVecFormModified_approx} can be compactly rewritten as
	\begin{eqnarray}
	\underbrace{\begin{bmatrix}\text{diag}(\mathbf{r}_{\rm x}) & - {\bf A}_{{\rm co}\mathbb{D}} \end{bmatrix}}_{\mathbf{G}_{\rm co}} \underbrace{\begin{bmatrix}\text{vec}({\bf C}) \\ {\boldsymbol \sigma}_{\rm z}\end{bmatrix}}_{{\boldsymbol \gamma}_{\rm co}} &=& {\boldsymbol \sigma}_{\rm n}.
	\label{eq:vec_covar_matrix_simplification}
	\end{eqnarray}
	The above system is underdetermined with $Q^2+D$ unknowns in $Q^{2}$ equations (note that some equations might even be redundant). However, as $\text{vec}({\bf C})$ has a Kronecker structure, the actual number of unknowns reduces to $Q$ and $\boldsymbol{\sigma}_{\rm z}$ is a sparse vector with non-zero elements at the location of the source DOAs. By considering the aforementioned constraints, the estimation problem can be cast as
	\begin{equation}
	\begin{aligned}
	&\underset{{\bf C},\, {\boldsymbol \sigma_{\rm z}}}{\rm min} \quad   \|\mathbf{G}_{\rm co}{\boldsymbol \gamma}_{\rm co} - {\boldsymbol \sigma}_{\rm n}\|_{2}^{2} + \lambda \|{\boldsymbol \sigma_{\rm z}}\|_0 \quad \text{s.t.} \quad ({\bf C},\, {\boldsymbol \sigma_{\rm z}}) \, \in \, \mathcal{C}_{\rm co}\\
	\end{aligned}
	\label{eq:AVS_Msensor_Nsource_multi_snapshot_scenario_sparse_model_opt_problem}
	\end{equation}
	where ${\boldsymbol \gamma}_{\rm co} = [{\rm vec}^T({\bf C}), \, {\boldsymbol \sigma}_{\rm z}^T]^T$, $\lambda$ is the regularization parameter, $\mathcal{C}_{\rm co} = \{({\bf C},\, {\boldsymbol \sigma_{\rm z}}) \, |\, {\boldsymbol \sigma}_{\rm z} \succeq {\bf 0}, \,{\mathbf{C}} = {\bf c}{\bf c}^H, \, c_1=c_2=1\}$ for APS arrays and $\mathcal{C}_{\rm co} = \{({\bf C},\, {\boldsymbol \sigma_{\rm z}}) \, |\, {\boldsymbol \sigma}_{\rm z} \succeq {\bf 0},\,{\mathbf{C}} = {\bf c}{\bf c}^H,\, c_1=1\}$ for AVS arrays. The optimization problem in~\eqref{eq:AVS_Msensor_Nsource_multi_snapshot_scenario_sparse_model_opt_problem} is non-convex due to the $l_{0}$ norm (cardinality) constraint and the rank-one equality constraint on ${\bf C}$. We can relax~\eqref{eq:AVS_Msensor_Nsource_multi_snapshot_scenario_sparse_model_opt_problem} by replacing the cardinality constraint with its convex approximation $\|{\boldsymbol \sigma}_{\rm z}\|_{\rm 1}$ and by replacing the rank-one equality constraint~(i.e., ${\mathbf{C}} = {\bf c}{\bf c}^H$) in the set $\mathcal{C}_{\rm co}$ with a convex inequality constraint~(i.e., ${\mathbf{C}} \succeq {\bf c}{\bf c}^H$). The new set which is same as $\mathcal{C}_{\rm co}$ except for the rank-one convex inequality constraint is denoted as $\tilde{\mathcal{C}}_{\rm co}$. The relaxed optimization problem can be expressed as,
	\begin{equation}
		\begin{aligned}
		&\underset{{\bf C},\, {\boldsymbol \sigma_{\rm z}}}{\rm min} \quad   \|\mathbf{G}_{\rm co}{\boldsymbol \gamma}_{\rm co} - {\boldsymbol \sigma}_{\rm n}\|_{2}^{2}+ \lambda \|{\boldsymbol \sigma}_{\rm z}\|_1 \quad \text{s.t.} \quad ({\bf C},\, {\boldsymbol \sigma_{\rm z}}) \, \in \, \tilde{\mathcal{C}}_{\rm co}.
		\end{aligned}
		\label{eq:AVS_Msensor_Nsource_multi_snapshot_scenario_sparse_model_opt_problem_convex}
	\end{equation}
	The convex inequality constraint, ${\bf C} \succeq {\bf c}{\bf c}^{H}$, is equivalent to $\begin{bmatrix}{\mathbf{C}} & \mathbf{c} \\ \mathbf{c}^{H} & 1 \end{bmatrix} \succeq 0$ from Schur's lemma.
	The resulting problem is a semi-definite programming problem that can be solved with any off-the-shelf solver. In practice, for the finite snapshot scenario, ${\bf C}$ obtained after solving~\eqref{eq:AVS_Msensor_Nsource_multi_snapshot_scenario_sparse_model_opt_problem_convex} might not be rank one and the closest estimates of the calibration parameters can be obtained from the first dominant singular vector of ${\bf C}$. However, when fewer number of snapshots are available, the sample covariance matrix will deviate from the assumed model and the rank-one relaxation in~\eqref{eq:AVS_Msensor_Nsource_multi_snapshot_scenario_sparse_model_opt_problem_convex} might not be tight. To further promote low rankness, we can introduce a trace constraint on ${\bf C}$ (which is the best convex relaxation of the rank constraint) in the cost function of~\eqref{eq:AVS_Msensor_Nsource_multi_snapshot_scenario_sparse_model_opt_problem_convex} as
	\begin{equation}
	\begin{aligned}
		&\underset{{\bf C},\, {\boldsymbol \sigma_{\rm z}}}{\rm min} \quad   \|\mathbf{G}_{\rm co}{\boldsymbol 	\gamma}_{\rm co} - {\boldsymbol \sigma}_{\rm n}\|_{2}^{2} + \lambda \|{\boldsymbol \sigma}_{\rm z}\|_1 + \beta \,\text{trace}\left({\bf C}\right) \\& \quad \text{s.t.} \quad ({\bf C},\, {\boldsymbol \sigma_{\rm z}}) \, \in \, \tilde{\mathcal{C}}_{\rm co},  \\
	\end{aligned}
	\label{eq:AVS_Msensor_Nsource_sparse_model_opt_problem_convex_trace_constraint}
	\end{equation}
	where $\beta > 0$ is the regularization parameter that allows for a trade off between the fit of the given data with respect to the assumed model and the trace constraint on ${\bf C}$. The formulation in~\eqref{eq:AVS_Msensor_Nsource_multi_snapshot_scenario_sparse_model_opt_problem_convex} can also be extended to sparse arrays for estimating DOAs (when there are more sources than sensors) and calibration parameters jointly as presented in~\cite{ramamohan2019blind}.
	
	Before ending this section, we remark that the proposed algorithms in Section~\ref{sec:calib_elementSpace}~and~Section~\ref{sec:calib_coarraySpace} are also useful for non-linear arrays using the identifiability conditions provided as a remark at the end of Section~\ref{sec:ambi_iden}.

	\section{Numerical experiments} \label{sec:NumericalExperiments}
	In this section, we present the numerical simulations to illustrate the performance of all the proposed solvers for the joint estimation of the source DOAs and calibration parameters. Firstly we consider the element-space model based solver in~\eqref{eq:DetModel} only for AVS linear arrays, as it does not require the presence of a reference source with known DOA as for APS linear arrays (see Section~\ref{sec:ambi_iden}). Then the covariance model in~\eqref{eq:AVS_Msensor_Nsource_multi_snapshot_scenario_sparse_model_opt_problem_convex} is considered for both the APS and AVS linear array.  Finally, we analyze the root mean square error~(RMSE) of the DOA estimates obtained from the presented algorithms and compare them with existing calibration methods. 
		
	\subsection{Element-space model}
	We consider a scenario with $M = 8$ AVSs arranged in a uniform linear array~(ULA) configuration where the spacing between the consecutive sensors is half a wavelength of the considered narrowband source signals. Further, we consider a scenario with $N = 6$ narrowband far-field signals impinging on the array from distinct DOAs with an observation period consisting of $L = 50$ snapshots. The grid is chosen to be uniform between $[0^{\circ}\, 180^{\circ}]$ with $1^{\circ}$ resolution. Without loss of generality, we assume the first channel of the first AVS in the array as the reference channel whose gain is 1 and phase is $0^{\circ}$. The gain and phase errors are picked from a uniform distribution over the interval [-3; 3] dB and $[-20^{\circ}; 20^{\circ}]$, respectively.
	\begin{figure}
		\begin{minipage}[h]{1\textwidth}
			\centering
			\includegraphics[width=0.7\textwidth]{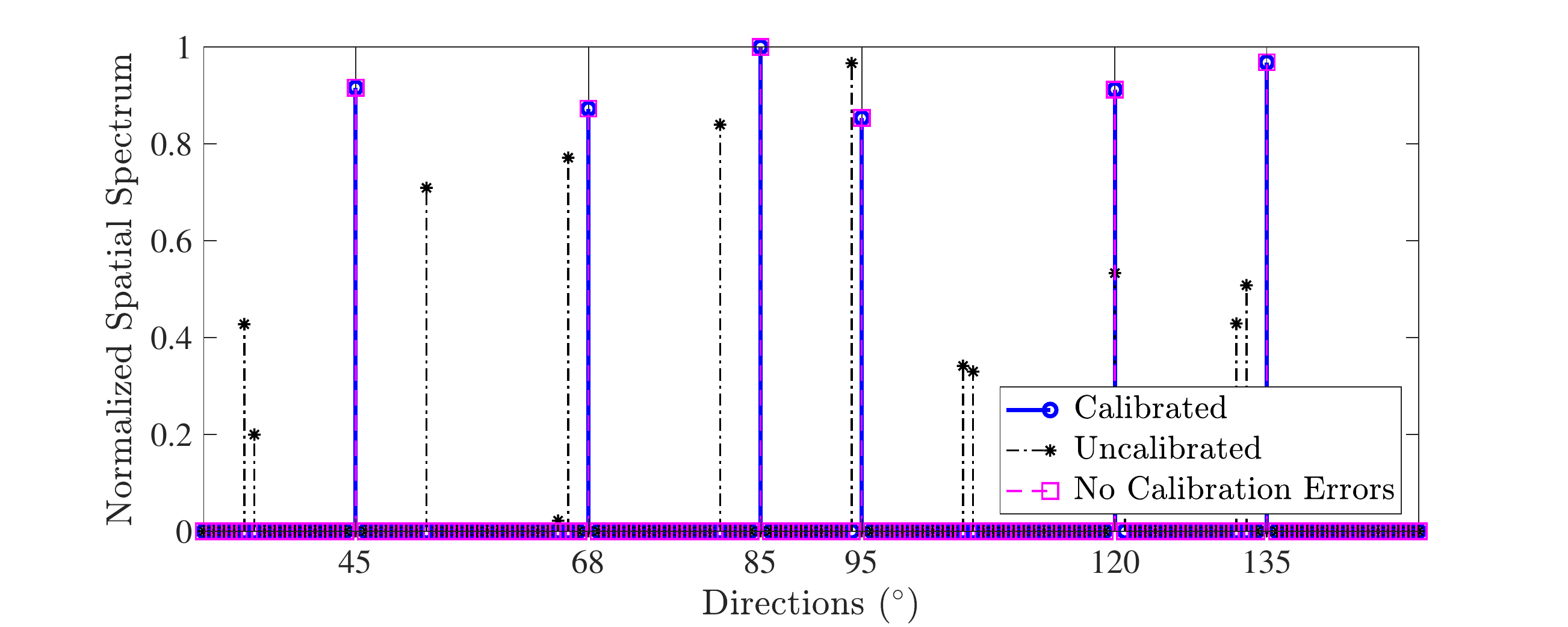}\\
			{\footnotesize{(a) $l_{1}$-SVD spectra without measurement noise.}}
		\end{minipage}
		\begin{minipage}[h]{1\textwidth}
			\centering
			\includegraphics[width=0.7\textwidth]{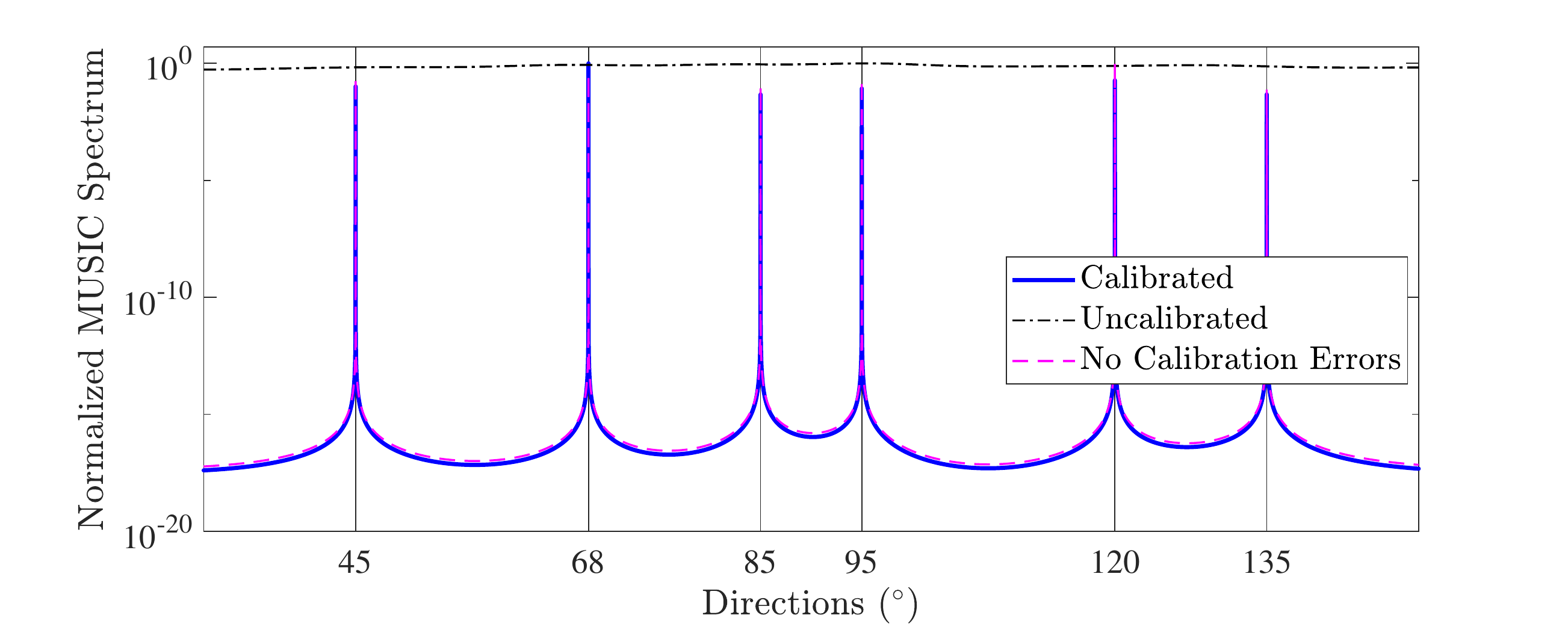}\\
			{\footnotesize{(b) MUSIC spectra without measurement noise.}}
		\end{minipage}  
		\begin{minipage}[h]{1\textwidth}
			\centering
			\includegraphics[width=0.7\textwidth]{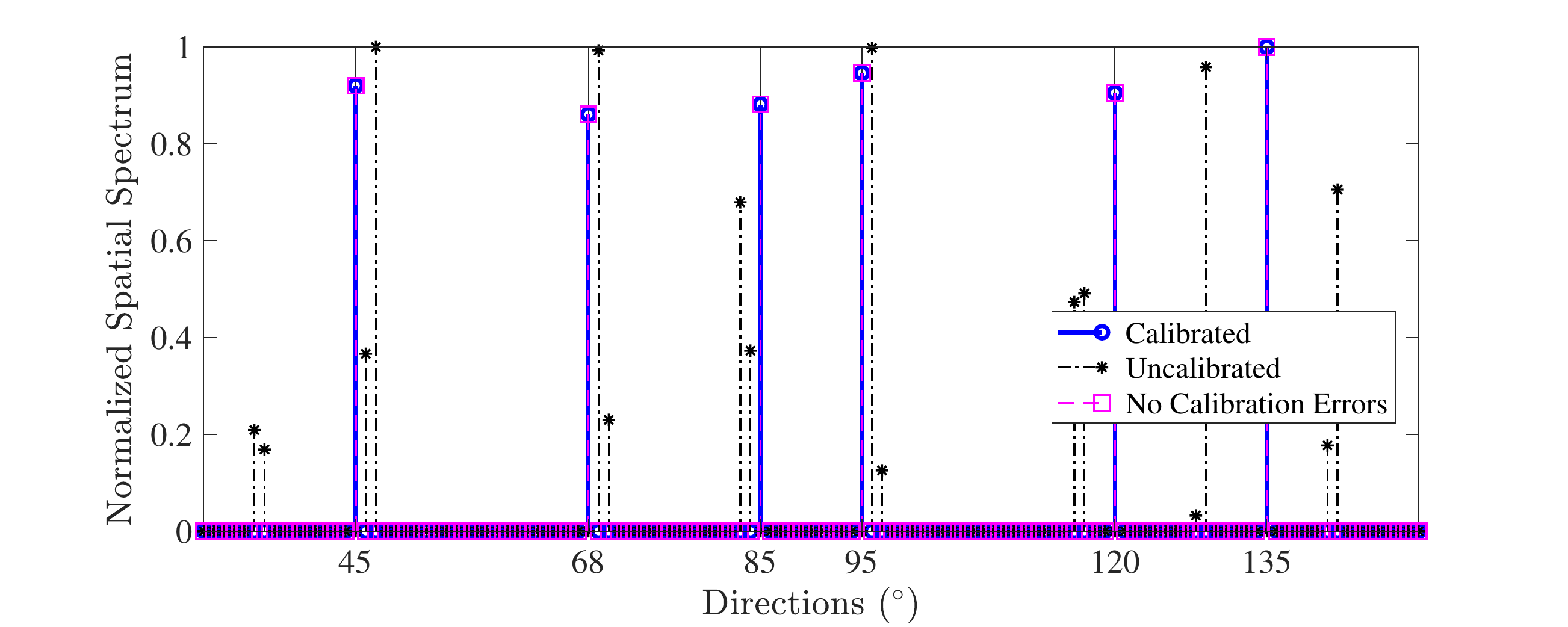}\\
			{\footnotesize{(c) $l_{1}$-SVD spectra with SNR = 10 dB.}}
		\end{minipage}
		\begin{minipage}[h]{1\textwidth}
			\centering
			\includegraphics[width=0.7\textwidth]{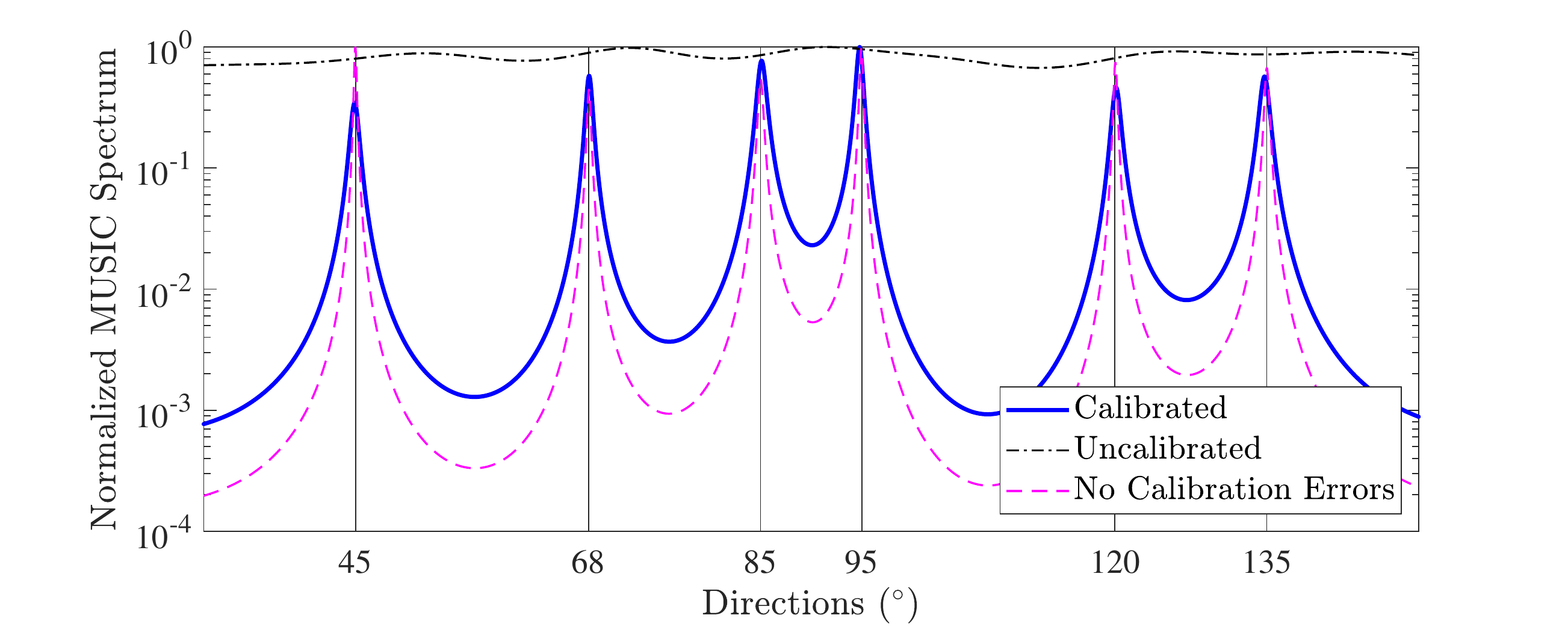}\\
			{\footnotesize{(d) MUSIC spectra with an SNR = 10 dB.}}
		\end{minipage}
		\caption{The $l_{1}$-SVD and MUSIC spectra using the element-space data model based solver in~\eqref{eq:DetModel} for an AVS ULA with $M = 8$, $N = 6$ and $L = 50$. The true DOAs are indicated by the black solid lines.}
		\label{fig:DetModel_Simulations}%
	\end{figure}	
	
	Based on the optimization problem in~\eqref{eq:DetModel}, the results of DOA estimation post calibration are presented in Fig.~\ref{fig:DetModel_Simulations}. For the choice of the regularization parameter $\lambda$, we follow the discrepancy principle as discussed in~\cite{malioutov2005sparse}, such that the residuals of the solution obtained using $\lambda$ match some known statistics of the noise. In order to verify the correctness of the formulation in~\eqref{eq:DetModel}, we initially considered an ideal scenario without measurement noise. The DOA spectra based on~\eqref{eq:DetModel} are presented in Fig.~\ref{fig:DetModel_Simulations}(a). It is seen in Fig.~\ref{fig:DetModel_Simulations}(a), that we recover the exact source DOAs after solving~\eqref{eq:DetModel}, where as for the uncalibrated data, the source DOA estimates based on the $l_{1}$-SVD algorithm~\cite{malioutov2005sparse} are very poor. Further, we considered the measurement data with a signal-to-noise ratio (SNR) of 10 dB and the corresponding DOA spectra obtained from solving~\eqref{eq:DetModel} are presented in Fig.~\ref{fig:DetModel_Simulations}(c), where we draw a similar inference as in Fig.~\ref{fig:DetModel_Simulations}(a).
	
	On the other hand, the issues of a pre-defined grid on the DOA estimates obtained after solving~\eqref{eq:DetModel} can be minimized by applying the MUSIC algorithm on the gain and phase compensated covariance matrix. The gain and phase errors are estimated from~\eqref{eq:DetModel}, and the corresponding MUSIC spectra are presented in Fig.~\ref{fig:DetModel_Simulations}(b). Specifically, in this two-step procedure, we only consider the estimates of the gain and phase errors to calibrate the array. Then we use the traditional MUSIC algorithm for DOA estimation. It can be inferred that for the ideal case without measurement noise, MUSIC with the uncalibrated data results in poor estimates, whereas the DOA estimates after calibration are in agreement with MUSIC obtained for the scenario without any calibration errors. Even for the measurement data with an SNR of 10 dB, MUSIC based on the calibrated data in Fig.~\ref{fig:DetModel_Simulations}(d) provides DOA estimates which are comparable with MUSIC without sensor errors.

	\subsection{Co-array data model}
	To illustrate the effectiveness of the covariance domain formulation provided in~\eqref{eq:AVS_Msensor_Nsource_multi_snapshot_scenario_sparse_model_opt_problem_convex} for the joint estimation of DOAs as well as the calibration parameters, we consider both a conventional uniform linear array~(ULA) with less sources than sensors and a sparse linear array with more sources than sensors. Here, all the far-field source DOAs are chosen to be on the grid and for the choice of the regularization parameter we follow the same approach as in the element-space approach. In both the scenarios, without loss of generality, for the APS arrays we considered the first two sensors as references whereas for the AVS arrays the first channel is considered as a reference with gain of 1 and phase of $0^{\circ}$. So for the APS arrays we count on the redundancies in the co-array to have a unique solution for the calibration parameters and DOAs as discussed at the end of Section~\ref{ref:coarrayIden}.	
		
	\hspace{10mm}
	\subsubsection{Uniform linear array with less sources than sensors}
\begin{figure}[h!]
		\begin{minipage}[h]{1\linewidth}
			\centering			
			\includegraphics[width=0.7\columnwidth]{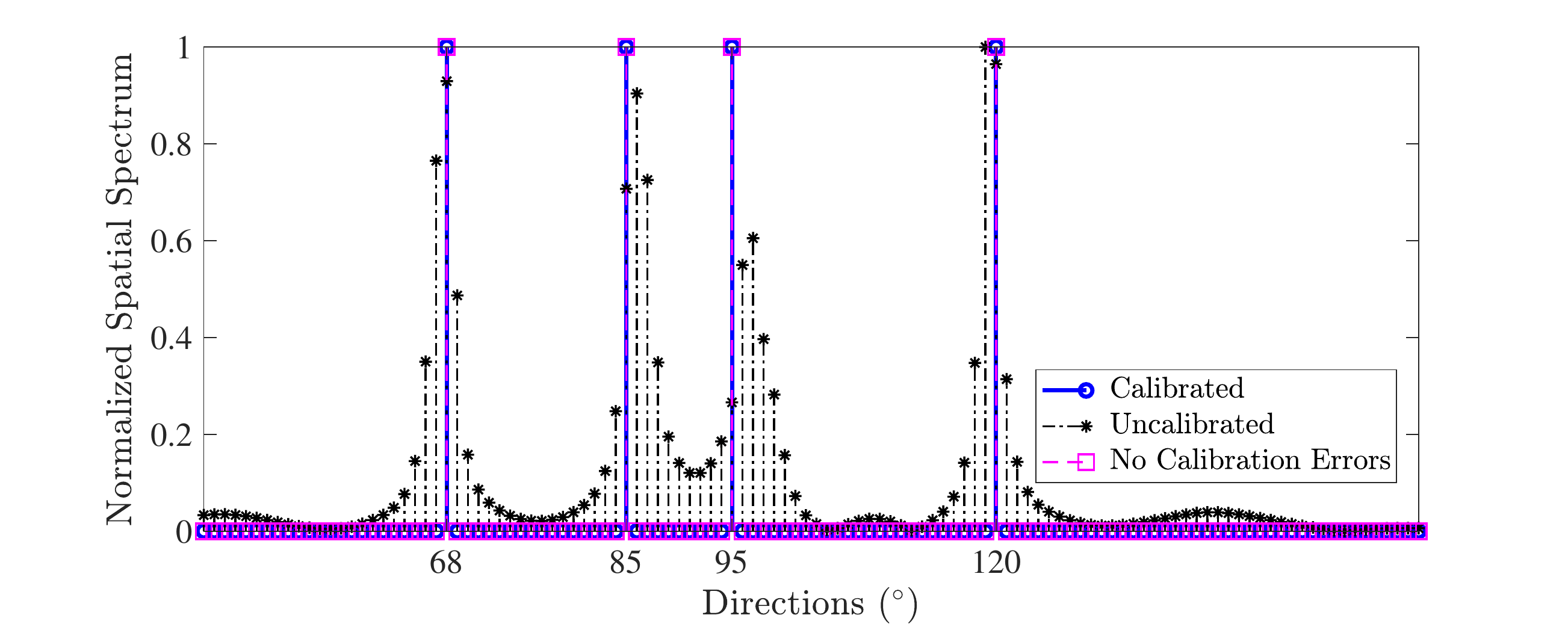}\\		
			{\footnotesize{(a) APS ULA - $L = \infty$.}}
		\end{minipage} \hfill  
		\begin{minipage}[h]{1\columnwidth}
			\centering			
			\includegraphics[width=0.7\linewidth]{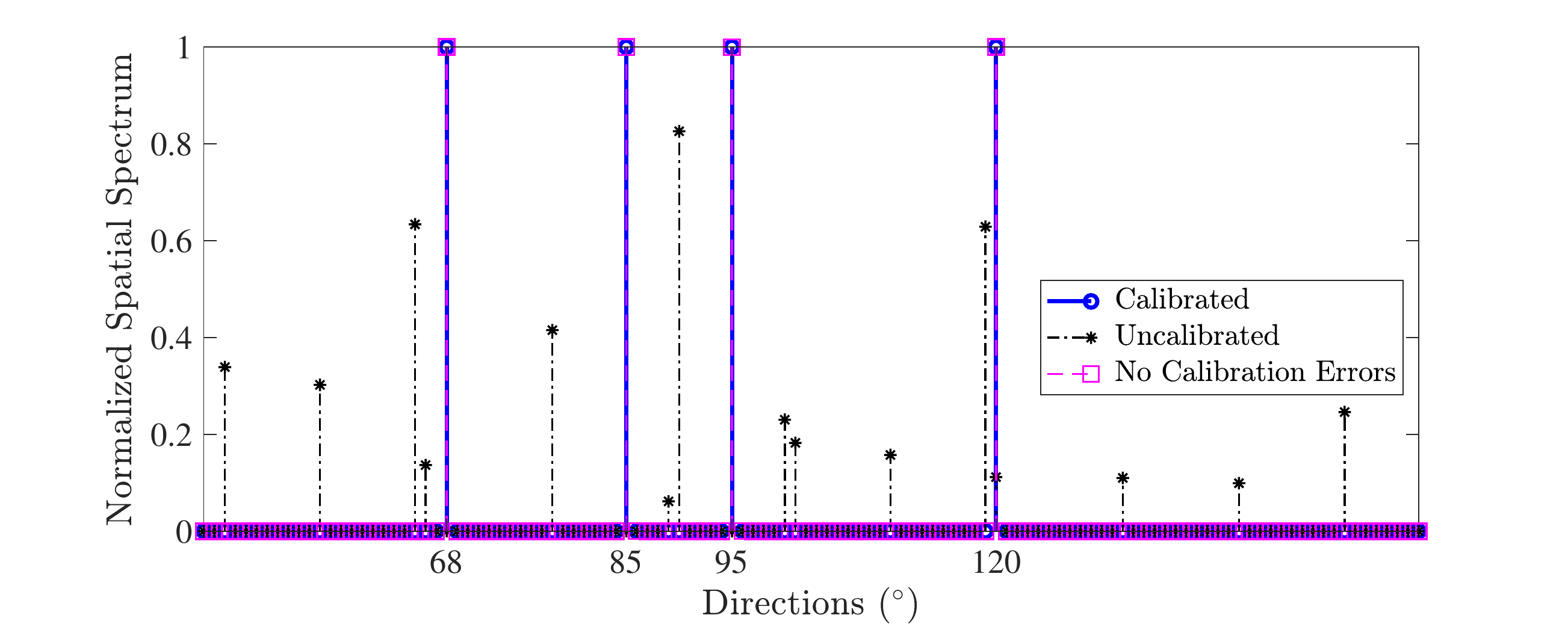}\\	
			{\footnotesize{(b) AVS ULA - $L = \infty$.}}
		\end{minipage} \\  
		\begin{minipage}[h]{1\columnwidth}
			\centering
			\includegraphics[width=0.7\linewidth]{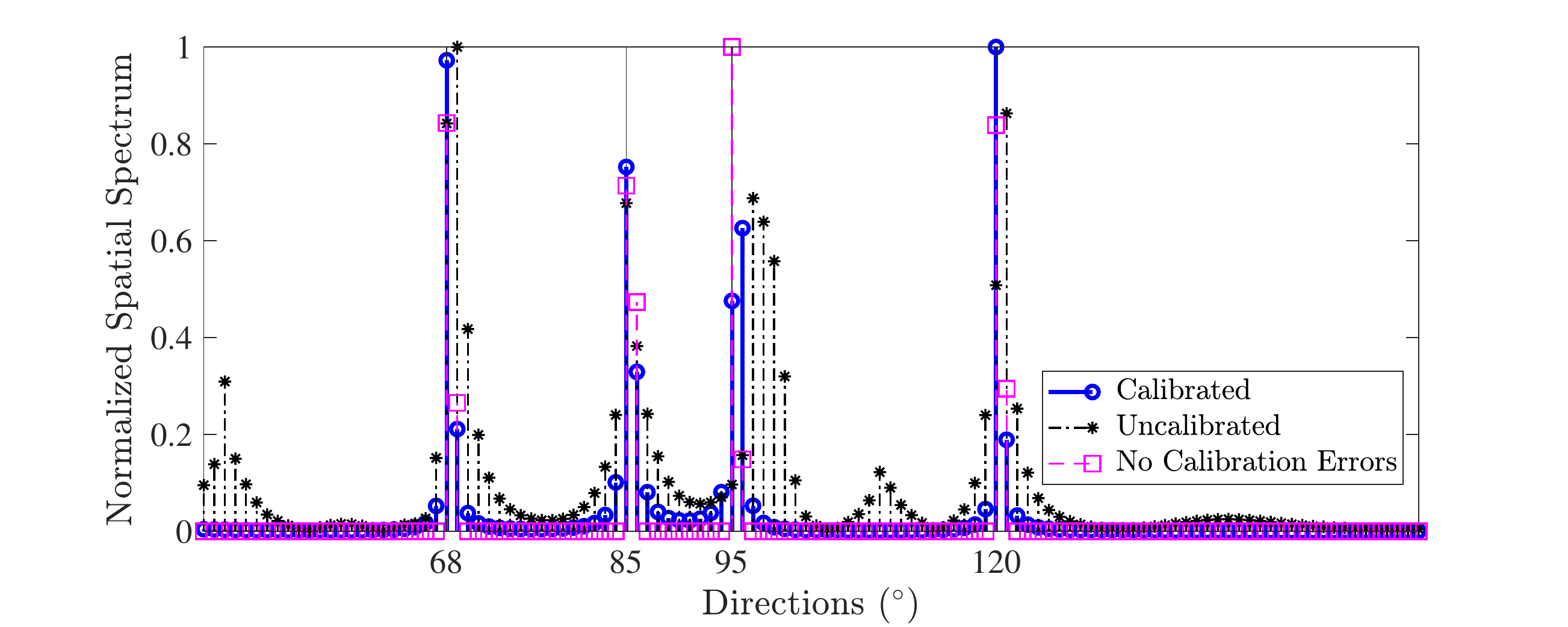}\\			
			{\footnotesize{(c) APS ULA - $L = 1000$.}}
		\end{minipage} \hfill  
		\begin{minipage}[h]{1\columnwidth}
			\centering
			\includegraphics[width=0.7\linewidth]{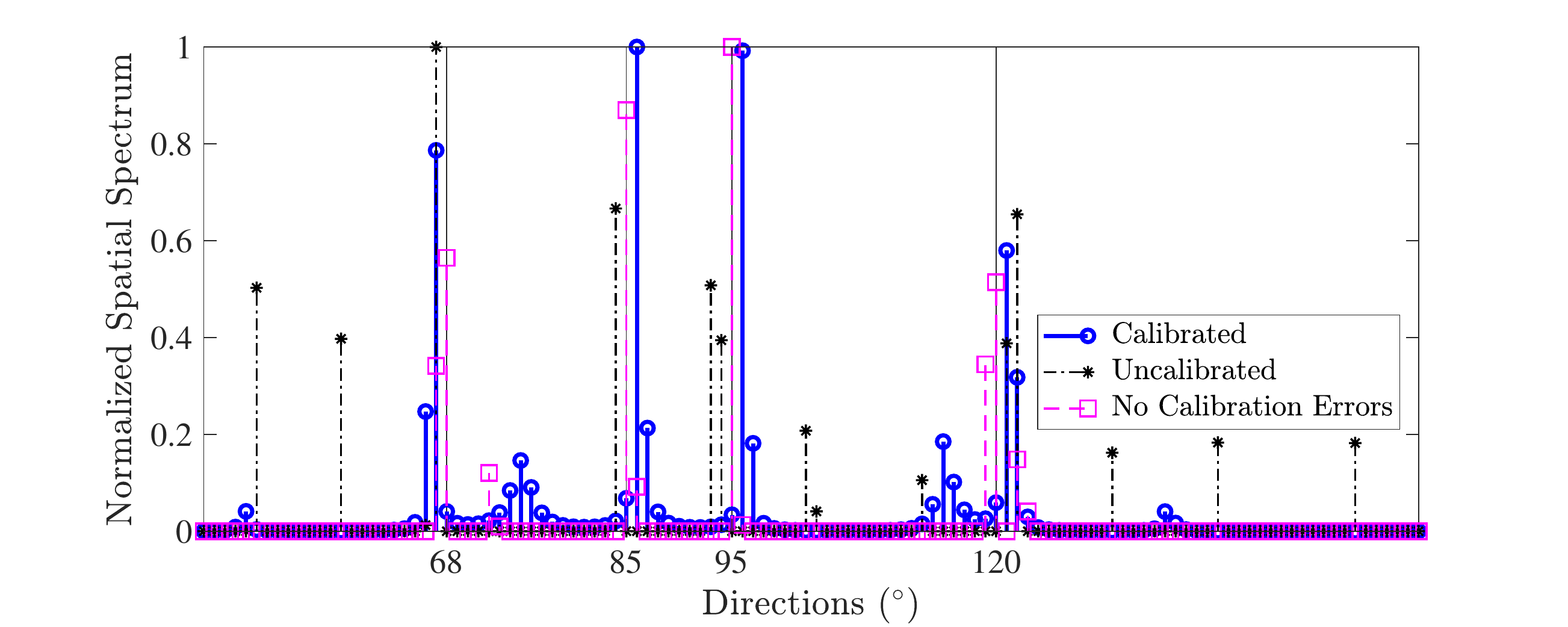}\\			
			{\footnotesize{(d) AVS ULA - $L = 1000$.}}
		\end{minipage}
		\caption{The $l_1$ norm based DOA spectra for both the APS and AVS ULA based on the co-array data model solver in~\eqref{eq:AVS_Msensor_Nsource_multi_snapshot_scenario_sparse_model_opt_problem_convex} with SNR = 10 dB, $M = 8$ and $N = 4$ far-field sources. The true DOAs are indicated by the black solid lines.}
		\label{fig:CovarModelULA_Simulations_StemPlots}	
	\end{figure}

	Consider a uniform linear array~(ULA) with $M = 8$, $N = 4$ far-field sources and SNR = 10 dB. Firstly, we will consider an ideal scenario with infinite snapshots, where $l_1$ norm based DOA spectra upon solving~\eqref{eq:AVS_Msensor_Nsource_multi_snapshot_scenario_sparse_model_opt_problem_convex} are plotted in Fig.~\ref{fig:CovarModelULA_Simulations_StemPlots}(a) for the APS ULA and in Fig.~\ref{fig:CovarModelULA_Simulations_StemPlots}(b) for the AVS ULA. It is seen that we exactly recover the source DOAs for both the APS and AVS ULA, indicating the exactness of the convex relaxation seen in~\eqref{eq:AVS_Msensor_Nsource_multi_snapshot_scenario_sparse_model_opt_problem_convex}. The uncalibrated data results in low resolution DOA spectra and very poor DOA estimates. 
	 
	Further, we consider a finite sample scenario with the observation period consisting of $L = 1000$ snapshots whose $l_1$ norm based DOA spectra upon solving~\eqref{eq:AVS_Msensor_Nsource_multi_snapshot_scenario_sparse_model_opt_problem_convex} are plotted in Fig.~\ref{fig:CovarModelULA_Simulations_StemPlots}(c) for the APS ULA and in Fig.~\ref{fig:CovarModelULA_Simulations_StemPlots}(d) for the AVS ULA. In Fig.~\ref{fig:CovarModelULA_Simulations_StemPlots}(c), the DOA spectra upon solving~\eqref{eq:AVS_Msensor_Nsource_multi_snapshot_scenario_sparse_model_opt_problem_convex} show an improvement compared to the DOA spectra computed with the uncalibrated data. However, the resulting DOA spectra still have low resolution, as the model considered in~\eqref{eq:AVS_Msensor_Nsource_multi_snapshot_scenario_sparse_model_opt_problem_convex} is not exact due to the finite sample approximation of the covariance matrix estimation. On the other hand, in Fig.~\ref{fig:CovarModelULA_Simulations_StemPlots}(d), the DOA spectra based on~\eqref{eq:AVS_Msensor_Nsource_multi_snapshot_scenario_sparse_model_opt_problem_convex} are significantly superior with high resolution compared to the DOA spectra computed with the uncalibrated data. However upon closer observation, we can notice that the DOA estimates are slightly biased for a couple of sources and also there are some spurious peaks in the DOA spectra. It is observed that the model mismatches due to the finite sample approximation of the covariance matrix estimation, has higher impact on reducing the sparsity of the DOA spectra for the APS ULA in comparison to an equivalent AVS ULA.

\begin{figure}[h!]
		\begin{minipage}[h]{1\linewidth}
			\centering
			\includegraphics[width=0.7\columnwidth]{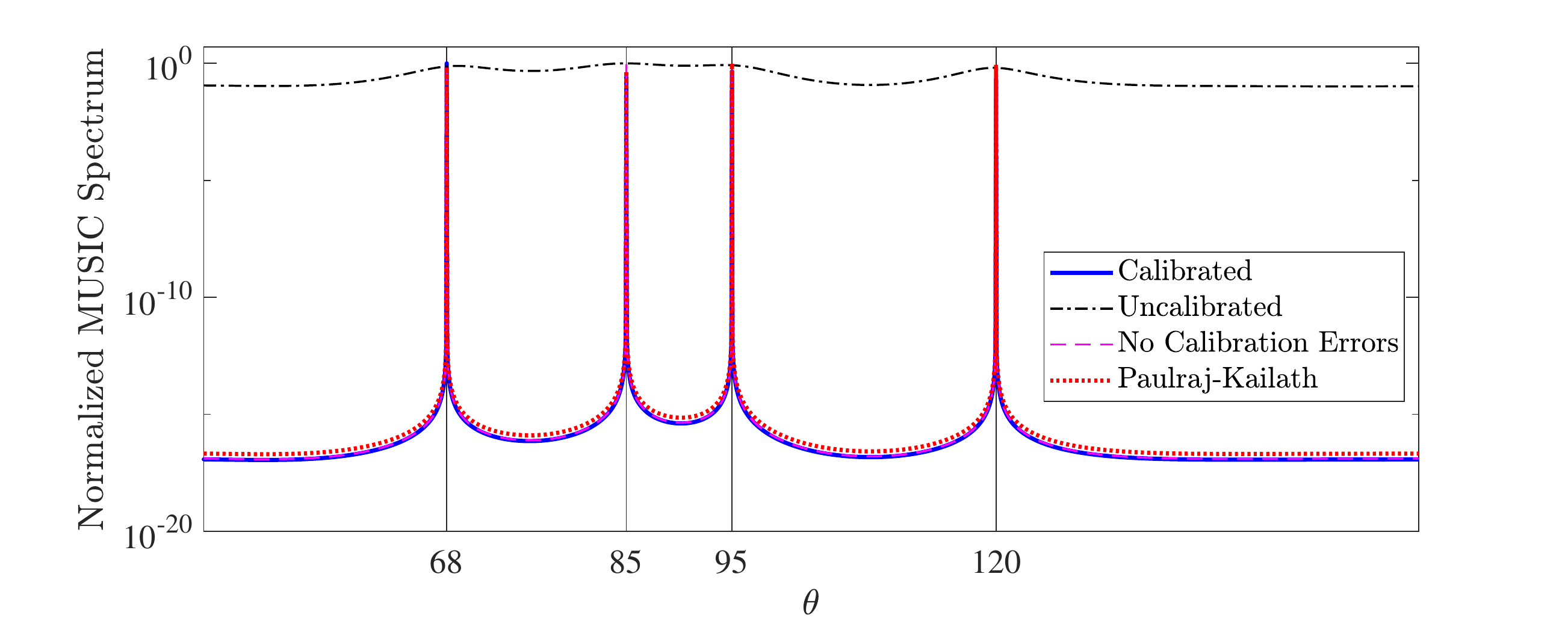}\\
			{\footnotesize{(a) APS ULA - $L = \infty$.}}
		\end{minipage} \hfill  
		\begin{minipage}[h]{1\columnwidth}
			\centering
			\includegraphics[width=0.7\linewidth]{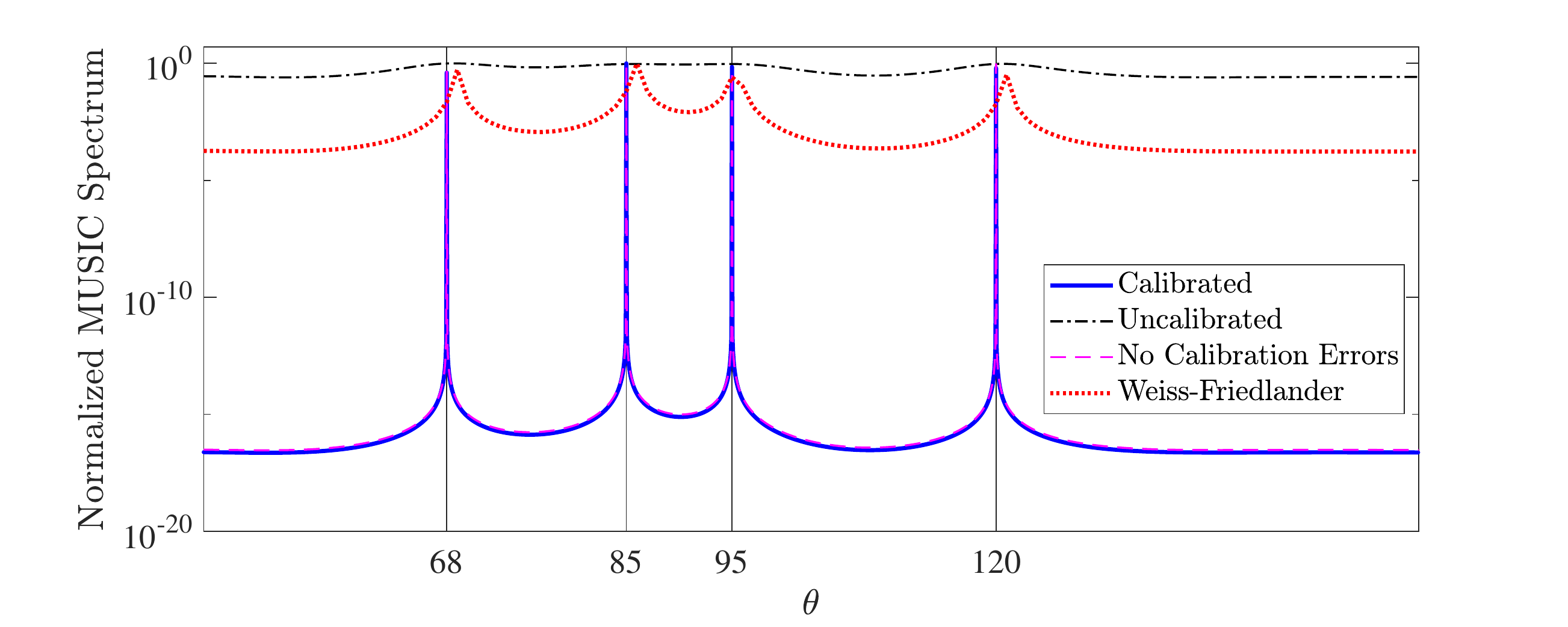}\\
			{\footnotesize{(b) AVS ULA - $L = \infty$.}}
		\end{minipage} \\  
		\begin{minipage}[h]{1\columnwidth}
			\centering
			\includegraphics[width=0.7\linewidth]{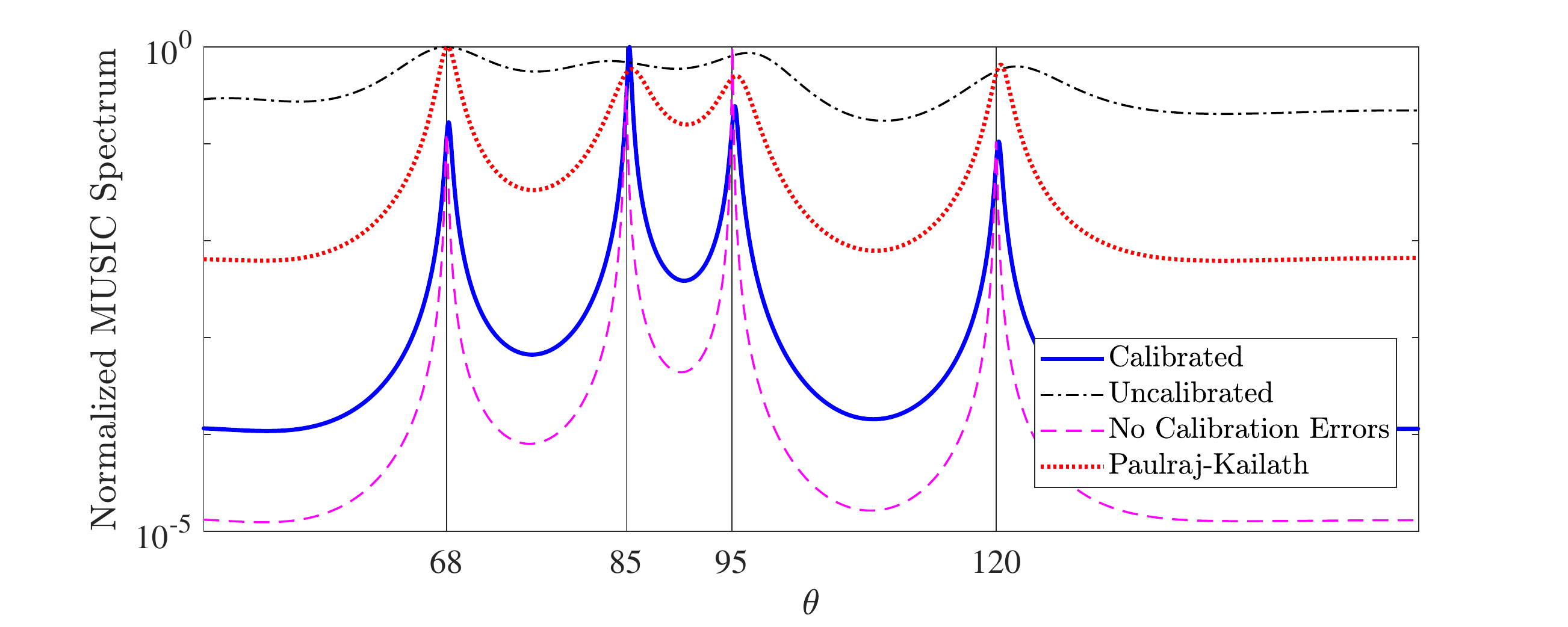}\\
			{\footnotesize{(c) APS ULA - $L = 1000$.}}
		\end{minipage} \hfill  
		\begin{minipage}[h]{1\columnwidth}
			\centering
			\includegraphics[width=0.7\linewidth]{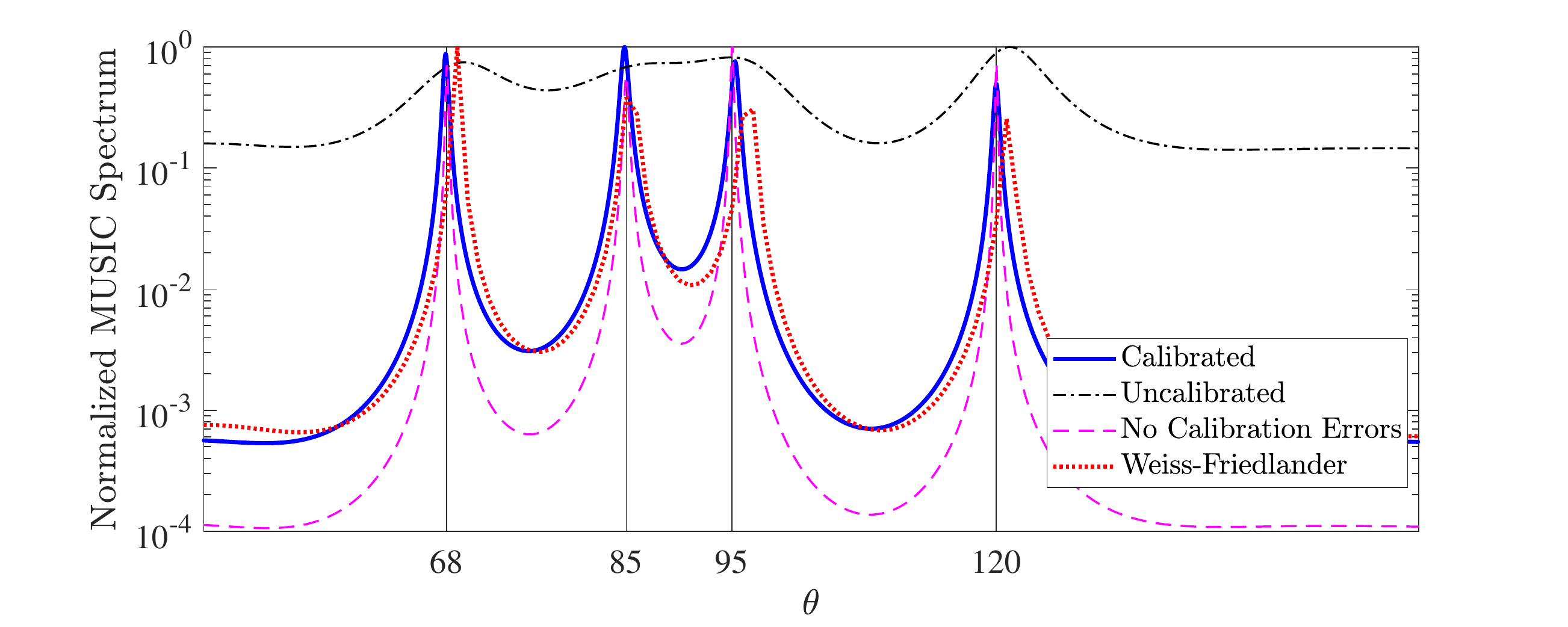}\\
			{\footnotesize{(d) AVS ULA - $L = 1000$.}}
		\end{minipage}
		\caption{The MUSIC spectra for both the APS and AVS ULA based on the co-array data model solver in~\eqref{eq:AVS_Msensor_Nsource_multi_snapshot_scenario_sparse_model_opt_problem_convex} with SNR = 10 dB, $M = 8$ and $N = 4$ far-field sources. The true DOAs are indicated by the black solid lines.}
		\label{fig:CovarModelULA_Simulations}	
	\end{figure}

	In order to overcome the discussed issues with DOA estimates and the effects of a predefined grid, similar to the element-space approach, a grid-free approach such as MUSIC algorithm can be applied on the measurement data in~\eqref{eq:covariance_matrix_yhat}, which is compensated for the gain and phase errors obtained from~\eqref{eq:AVS_Msensor_Nsource_multi_snapshot_scenario_sparse_model_opt_problem_convex}. Those MUSIC spectra based on the calibrated data are presented in Fig.~\ref{fig:CovarModelULA_Simulations}. The results in Fig.~\ref{fig:CovarModelULA_Simulations}(b) for the AVS ULA is compared with~\cite{weiss1990eigenstructure} (referred to as Weiss-Friedlander approach). The results in Fig.~\ref{fig:CovarModelULA_Simulations}(a) for the APS ULA is compared with~\cite{paulraj1985direction} (referred to as the Paulraj-Kailath approach\footnote{{During the submission of this manuscript it came to the authors' attention that an improved version of~\cite{paulraj1985direction} for scalar sensor arrays that considers an optimally-weighted least squares~(OWLS) approach was proposed in~\cite{weiss2020asymptotically}.}}), as the Weiss-Friedlander approach is not effective for linear scalar sensor arrays. 
	
	The MUSIC spectra for an ideal scenario with infinite snapshots are plotted in Fig.~\ref{fig:CovarModelULA_Simulations}(a) for the APS ULA and in Fig.~\ref{fig:CovarModelULA_Simulations}(b) for the AVS ULA. Similar to Figs.~\ref{fig:CovarModelULA_Simulations_StemPlots}(a)~and~(b), we obtain an optimal solution after calibration where the results are the same as those obtained from measurements with no calibration errors. 
	For the finite snapshot case, in Figs.~\ref{fig:CovarModelULA_Simulations}(c)~and~(d), we see that the MUSIC spectra have a higher resolution and improved estimates compared to the equivalent $l_1$ norm based DOA spectra. On contrary, the spectra based on the uncalibrated data is not able to resolve all the sources and the resolution of the spectra is also degraded. Further, for the APS ULA in Fig.~\ref{fig:CovarModelULA_Simulations}(c), the proposed approach outperforms~\cite{paulraj1985direction}, and for the AVS ULA in Fig.~\ref{fig:CovarModelULA_Simulations}(d), it can be observed that although~\cite{weiss1990eigenstructure} results in a smaller variance compared to the proposed approach, the estimates are highly biased.
	
	It can be summarized that based on the formulation in~\eqref{eq:AVS_Msensor_Nsource_multi_snapshot_scenario_sparse_model_opt_problem_convex}, it is possible to jointly estimate both the calibration errors as well as the source DOAs and the estimation results are good when the number of time snapshots are higher and the grid-mismatches are minimal. However, when the number of time snapshots are limited and we have a pre-defined grid, solving~\eqref{eq:AVS_Msensor_Nsource_multi_snapshot_scenario_sparse_model_opt_problem_convex} can be used as a pre-conditioning step to estimate the calibration errors. Then a grid-free approach such as MUSIC can be applied on the gain and phase errors compensated measurement data to obtain improved and reliable DOA estimates. 

	\hspace{10mm}
	\subsubsection{Sparse array with more sources than sensors}
\begin{figure}[h!]
		\begin{minipage}[h]{1\linewidth}
			\centering
			\includegraphics[width=0.7\columnwidth]{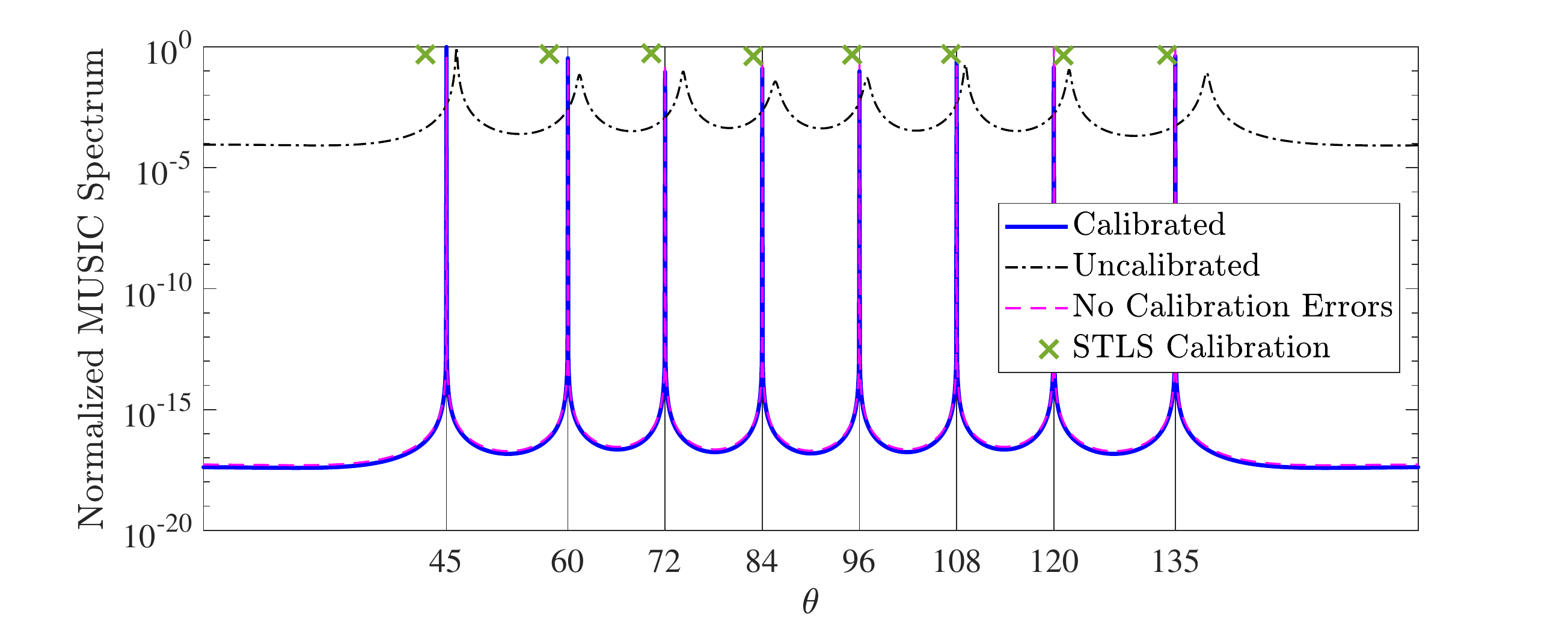}\\
			{\footnotesize{(a) APS sparse array - $L = \infty$.}}
		\end{minipage} \hfill  
		\begin{minipage}[h]{1\columnwidth}
			\centering
			\includegraphics[width=0.7\linewidth]{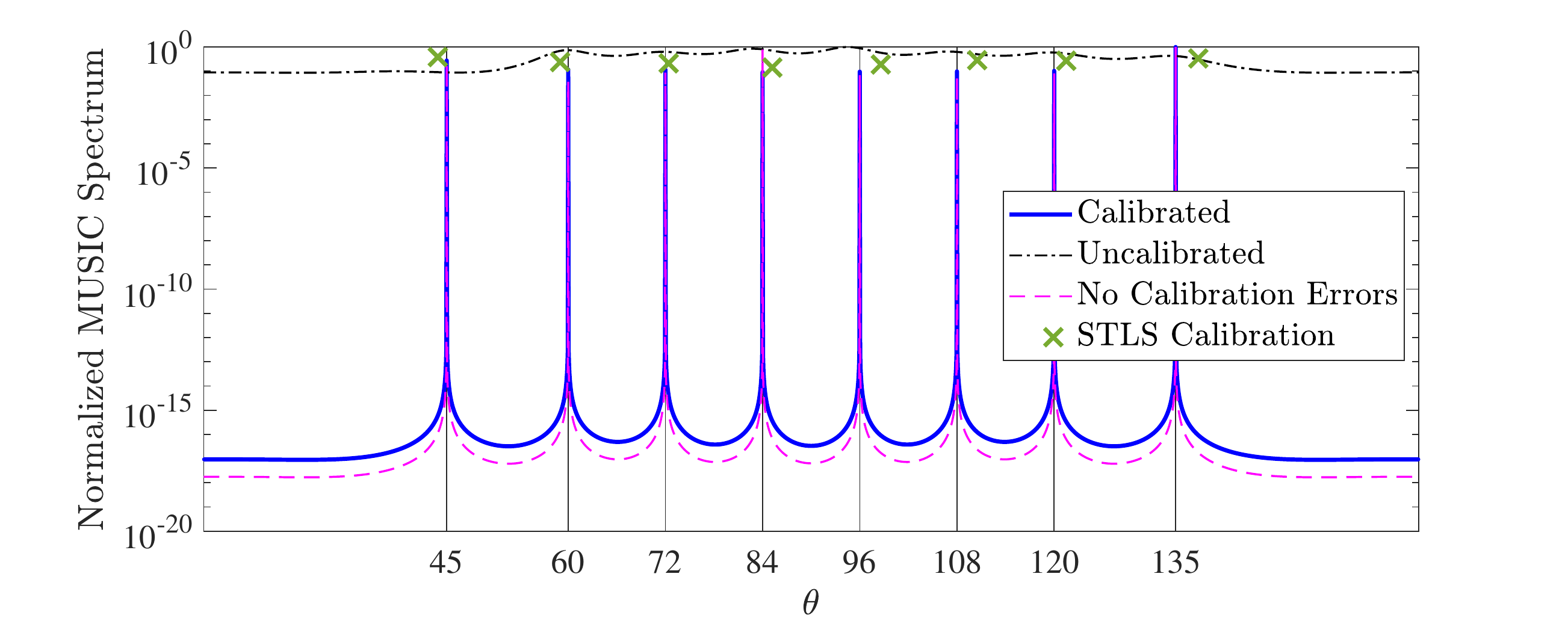}\\
			{\footnotesize{(b) AVS sparse array - $L = \infty$.}}
		\end{minipage} \\  
		\begin{minipage}[h]{1\columnwidth}
			\centering
			\includegraphics[width=0.7\linewidth]{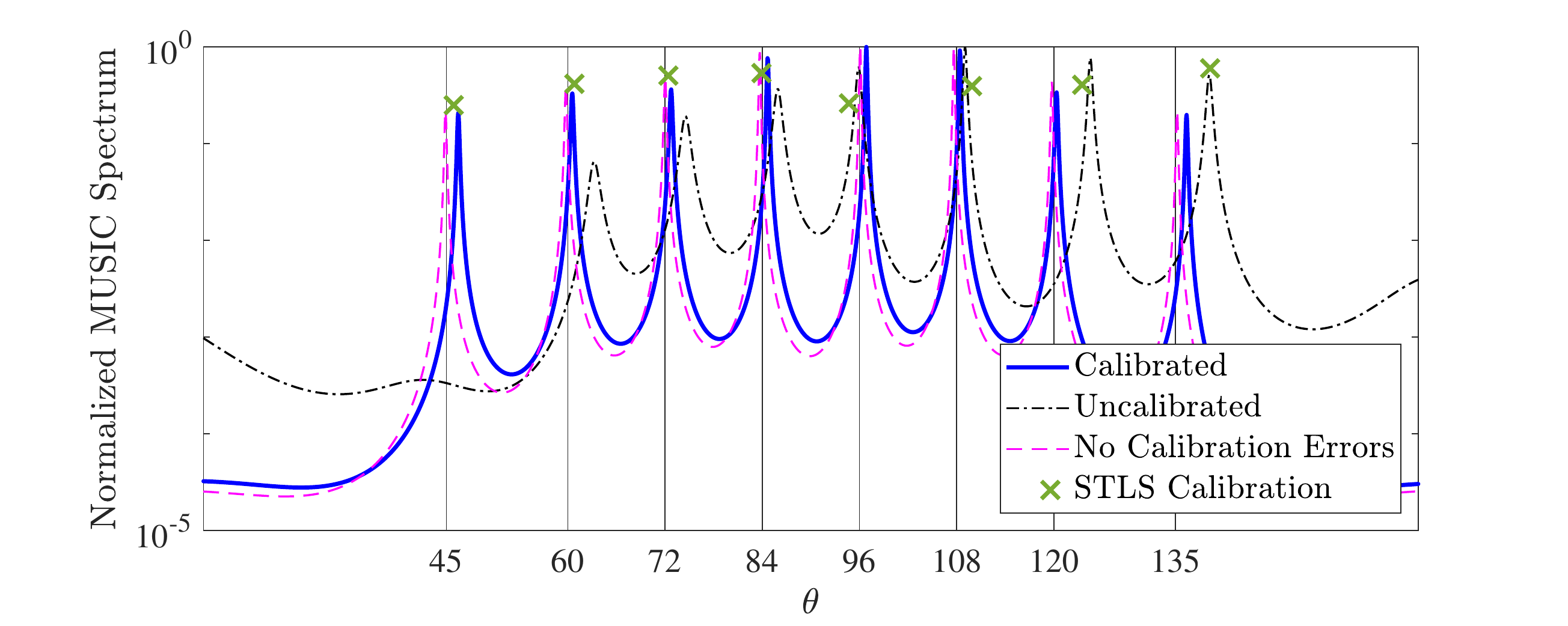}\\
			{\footnotesize{(c) APS sparse array - $L = 500$.}}
		\end{minipage} \hfill  
		\begin{minipage}[h]{1\columnwidth}
			\centering
			\includegraphics[width=0.7\linewidth]{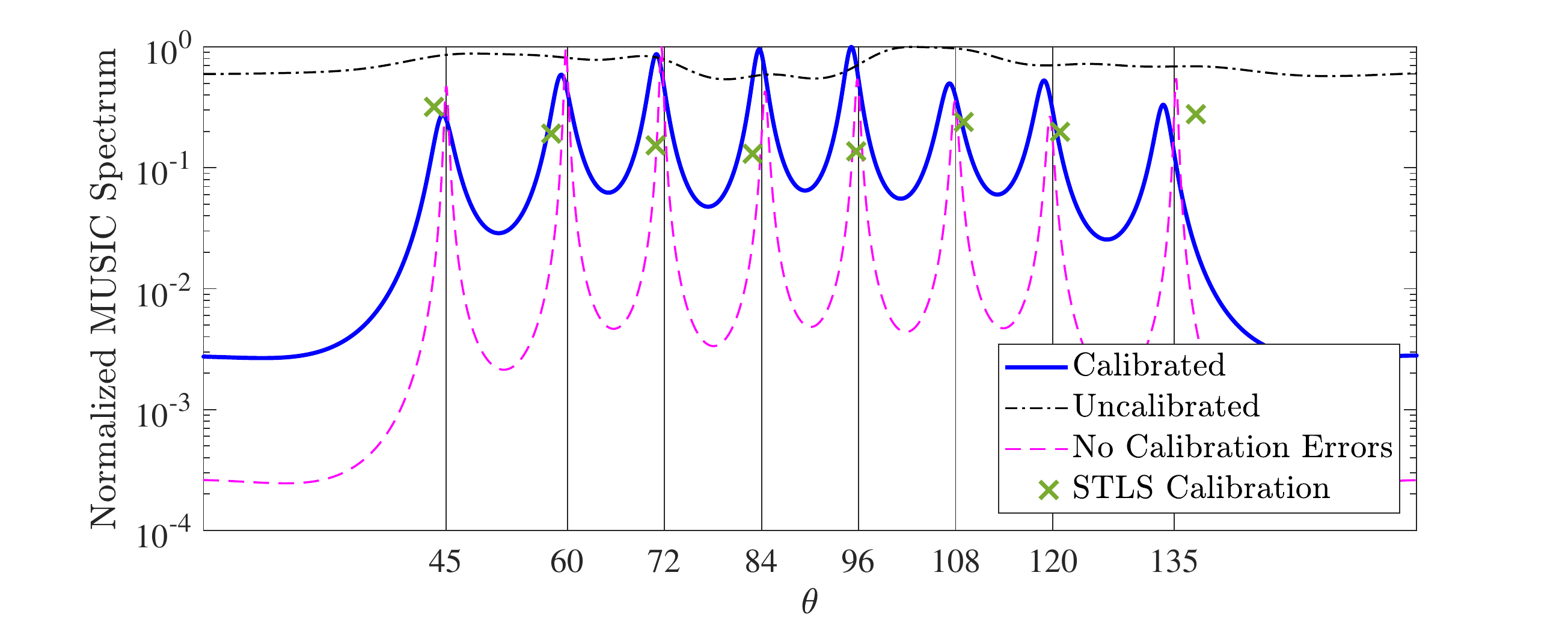}\\
			{\footnotesize{(d) AVS sparse array - $L = 500$.}}
		\end{minipage}
		\caption{The spatial smoothing MUSIC (SS MUSIC) spectra for both the APS and AVS sparse linear array based on the co-array data model solver in~\eqref{eq:AVS_Msensor_Nsource_multi_snapshot_scenario_sparse_model_opt_problem_convex} with SNR = 10 dB, $M = 6$, ${\bf p} = [0\, 1\, 2\, 3\, 6\, 9]^T$ and $N = 8$ far-field sources. The true DOAs are indicated by the black solid lines.}
		\label{fig:CovarModel_Simulations}	
	\end{figure}

	 Consider a hole-free sparse linear array with $M = 6$, ${\bf p} = [0\, 1\, 2\, 3\, 6\, 9]^T$, $N = 8$ far-field sources and SNR = 10 dB. The rank of the ${\bf G}$ matrix, [cf.~\eqref{eq:sysOfEqns_PhaseRelations}], for the considered sparse array is 4 (i.e., $M-2$). For this scenario we just present spatial smoothing MUSIC (SS MUSIC) spectra based on the gain and phase compensated measurement data. The calibration errors are estimated by evaluating the proposed formulation in~\eqref{eq:AVS_Msensor_Nsource_multi_snapshot_scenario_sparse_model_opt_problem_convex}.
	 Firstly, we will consider an ideal scenario with infinite snapshots, i.e., the exact covariance matrix as in~\eqref{eq:covariance_matrix_yhat} is considered, and the corresponding SS MUSIC spectra are plotted in Fig.~\ref{fig:CovarModel_Simulations}(a) for the APS array and in Fig.~\ref{fig:CovarModel_Simulations}(b) for the AVS array. Further, we consider a finite sample scenario with the observation period consisting of $L = 500$ snapshots whose SS MUSIC spectra are plotted in Fig.~\ref{fig:CovarModel_Simulations}(c) for the APS array and Fig.~\ref{fig:CovarModel_Simulations}(d) for the AVS array. The results of SS MUSIC for both the APS and AVS array are compared with the sparse total least squares~(STLS) calibration approach~\cite{han2015calibrating}.

	From Figs.~\ref{fig:CovarModel_Simulations}(a)~and~(b), it can be inferred that the resolution of the SS MUSIC spectra with uncalibrated data is poor whereas we obtain the true DOAs for both the APS and AVS arrays with calibrated data similar to the case for measurement data with no calibration errors. On the other hand, even for infinite data records, the STLS calibration approach~\cite{han2015calibrating} leads to a sub optimal solution, where source DOAs are not perfectly recovered.
	For the finite snapshot case, both in Figs.~\ref{fig:CovarModel_Simulations}(c)~and~(d), we see that post calibration, the SS MUSIC spectra have a higher resolution and are comparable to the scenario with no calibration errors, whereas the spectra based on the uncalibrated data are not able to resolve all the sources and the resolution of the spectra is also degraded. Furthermore, for both the APS and AVS sparse array with 500 snapshots, the performance of our proposed method is better than the STLS calibration approach~\cite{han2015calibrating}.

	\subsection{Monte-Carlo experiments} \label{subsec:MonteCarloSimulations}
	In this section we will study the statistical behavior through the root mean square error (RMSE) of the DOA estimation based on the proposed calibration procedure for different scenarios. For the analysis we consider both AVSs and APSs arranged in a uniform linear array~(ULA)
	%
	with $M = 8$ and three far-field sources, i.e., $N = 3$ and $\boldsymbol{\theta} = [78^{\circ}, 90^{\circ}, 102^{\circ}]$. The gain and phase perturbations follow a uniform distribution over the interval of [-2, 2] dB and $[-40^{\circ}, 40^{\circ}]$, respectively. For both the element-space formulation~\eqref{eq:DetModel} and covariance domain formulation~\eqref{eq:AVS_Msensor_Nsource_multi_snapshot_scenario_sparse_model_opt_problem_convex}, we have chosen the pre-defined grid between $0^{\circ}$ and $180^{\circ}$ with $1^{\circ}$ resolution. The RMSE of the DOA estimates based on the $l_1$ norm spectra (either by solving~\eqref{eq:DetModel}~or~\eqref{eq:AVS_Msensor_Nsource_multi_snapshot_scenario_sparse_model_opt_problem_convex}) as well as the MUSIC spectra are presented for the considered scenarios.
	
\begin{figure}[h!]
		\begin{minipage}[h]{1\linewidth}
			\centering
			\includegraphics[width=0.7\linewidth]{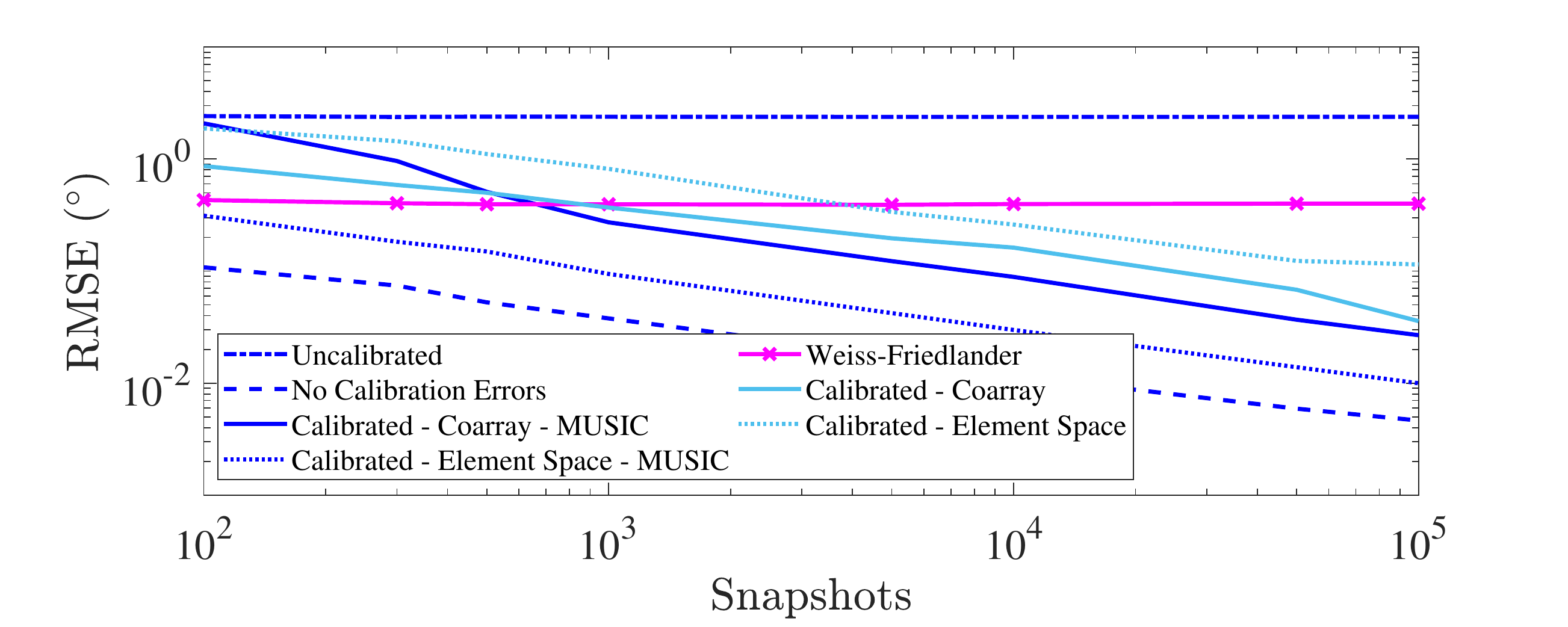}\\
			{\footnotesize{(a) AVS ULA with SNR = 10 dB.}}
		\end{minipage}
		\begin{minipage}[h]{1\linewidth}
			\centering
			\includegraphics[width=0.7\linewidth]{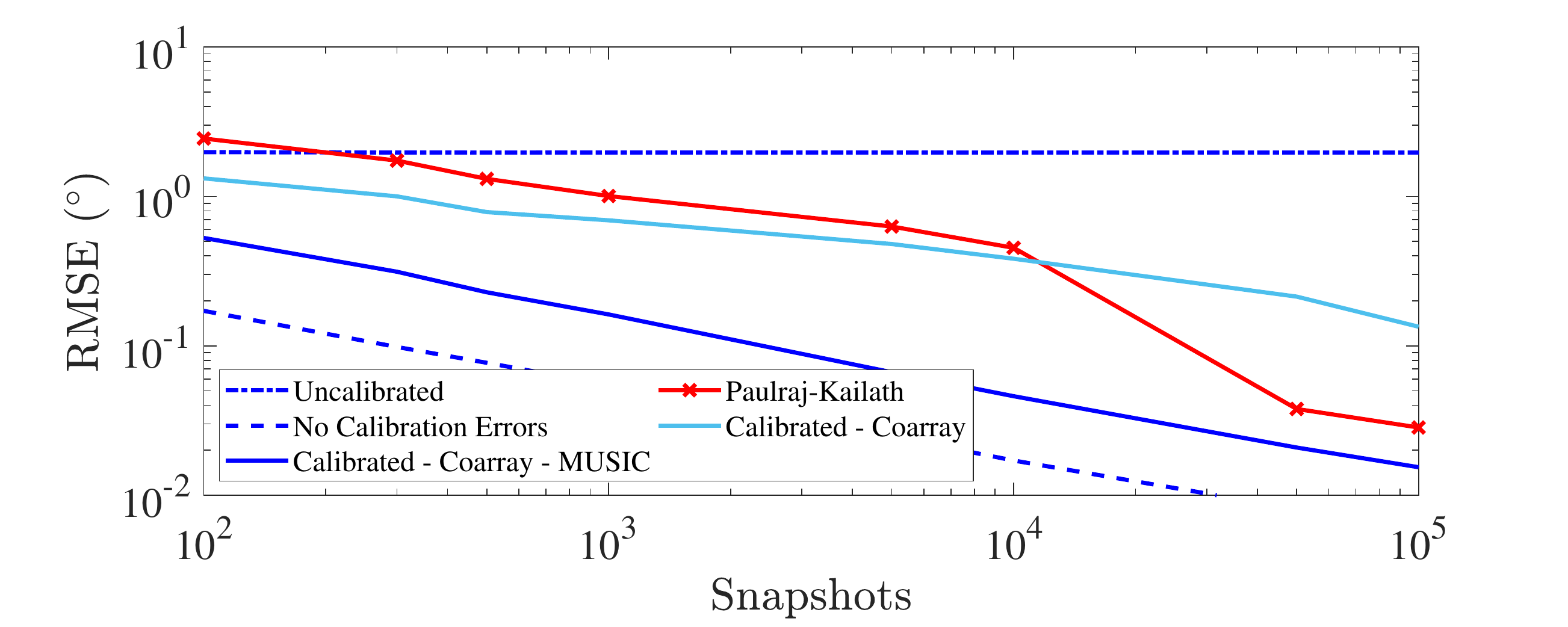}\\
			{\footnotesize{(b) APS ULA with SNR = 10 dB.}}
		\end{minipage} 		
		\caption{RMSE variation of the DOA estimates for the source at $90^{\circ}$ using both the APS and AVS ULA with $M = 8$, $N = 3$ and $\boldsymbol{\theta} = [78^{\circ}, 90^{\circ}, 102^{\circ}]$ for a fixed SNR as the number of snapshots are varying.}
		\label{fig:RMSE_Simulations_SnapshotsVarying}
	\end{figure}

	\paragraph*{Fixed SNR and varying snapshots}
	 The RMSE of the DOA estimates for the source present at $90^{\circ}$ based on 500 Monte-Carlo trials for both the APS and AVS ULA	are presented in Fig.~\ref{fig:RMSE_Simulations_SnapshotsVarying}. Here the calibration errors and SNR of 10 dB were fixed for all the trails while the number of snapshots are varying. 
	 The RMSE of the DOA estimates in Fig.~\ref{fig:RMSE_Simulations_SnapshotsVarying} based on the $l_1$ norm spectra by solving~\eqref{eq:DetModel} is referred to as "Calibrated - Element Space" and by solving~\eqref{eq:AVS_Msensor_Nsource_multi_snapshot_scenario_sparse_model_opt_problem_convex} is referred to as "Calibrated - Coarray". Further, the RMSE in the DOA estimates in Fig.~\ref{fig:RMSE_Simulations_SnapshotsVarying} based on the MUSIC spectra by solving~\eqref{eq:DetModel} is referred to as "Calibrated - Element Space - MUSIC" and by solving~\eqref{eq:AVS_Msensor_Nsource_multi_snapshot_scenario_sparse_model_opt_problem_convex}  is referred to as "Calibrated - Coarray - MUSIC".
	
	In Fig.~\ref{fig:RMSE_Simulations_SnapshotsVarying}(a), we considered the AVS ULA with an SNR of 10 dB. It is seen that as the number of snapshots increases, the RMSE of the DOA estimates for the uncalibrated case does not decrease, whereas after calibration based on both the $l_1$ norm spectra and the MUSIC spectra, the results approach the ideal scenario with no calibration errors. For a given number of snapshots, the MUSIC based DOA estimates result in lower RMSE values when compared with the equivalent $l_1$ norm based DOA estimates, further emphasizing the fact that the calibration estimates are robust to the model mismatches 
	while solving either~\eqref{eq:DetModel}~or~\eqref{eq:AVS_Msensor_Nsource_multi_snapshot_scenario_sparse_model_opt_problem_convex}. On the other hand, the RMSE of the DOA estimates based on the Weiss-Friedlander~(WF) approach is also presented in Fig.~\ref{fig:RMSE_Simulations_SnapshotsVarying}, where the calibration parameters were initialized with a gain of 1 and a phase of $0^{\circ}$. It is seen that the RMSE of the DOA estimates decreases initially, however it tends to saturate as the number of snapshots increases as it leads to a sub-optimal solution depending on the initialization.
	Also it can be observed that the DOA estimates based on the MUSIC spectra with calibration parameters estimated from~\eqref{eq:AVS_Msensor_Nsource_multi_snapshot_scenario_sparse_model_opt_problem_convex} require more snapshots to obtain better DOA estimates with low RMSE as the finite sample errors in the estimation of the covariance matrix are high for a low number of snapshots and those are not modeled in the formulation of~\eqref{eq:AVS_Msensor_Nsource_multi_snapshot_scenario_sparse_model_opt_problem_convex}. Furthermore, based on the MUSIC spectra in Fig.~\ref{fig:RMSE_Simulations_SnapshotsVarying}(a), it can be observed that the performance of the element-space approach is far superior than the covariance domain approach. 
	
\begin{figure}[h!]
		\begin{minipage}[h]{1\linewidth}
			\centering
			\includegraphics[width=0.7\linewidth]{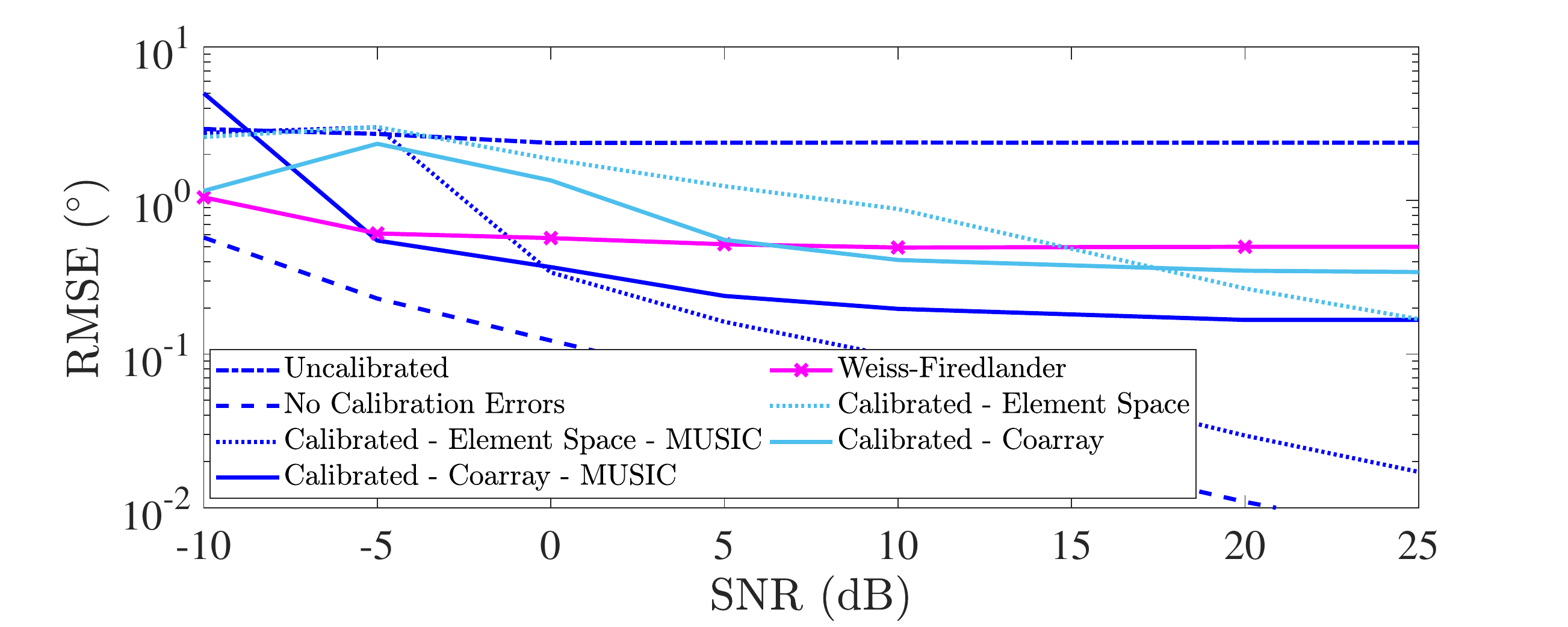}\\
			{\footnotesize{(a) AVS ULA - $L=1000$.}}
		\end{minipage}
		\begin{minipage}[h]{1\linewidth}
			\centering
			\includegraphics[width=0.7\linewidth]{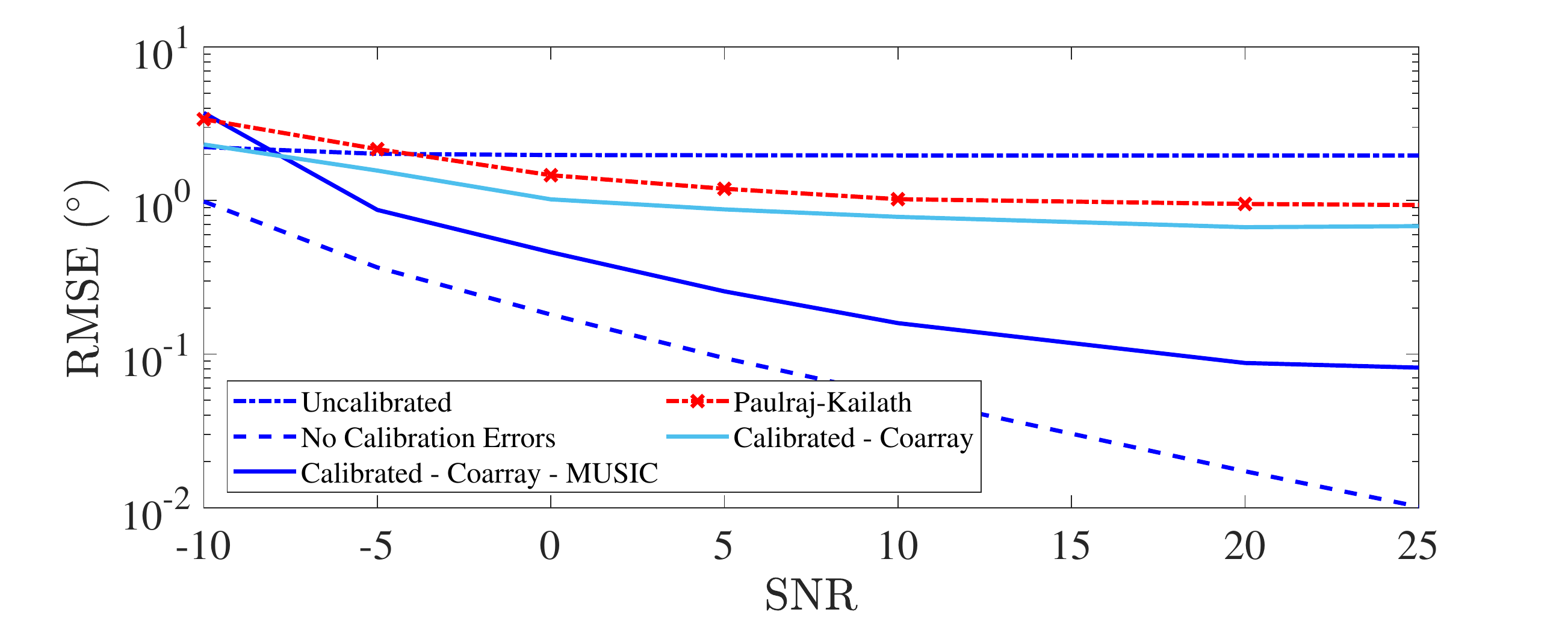}\\
			{\footnotesize{(b) APS ULA - $L = 1000$.}}
		\end{minipage} 		
		\caption{RMSE variation of the DOA estimates for the source at 90$^\circ$ using both the APS and AVS ULA with $M = 8$, $N = 3$ and $\boldsymbol{\theta} = [78^{\circ}, 90^{\circ}, 102^{\circ}]$ as the SNR varies for a fixed number of snapshots.}
		\label{fig:RMSE_Simulations_SNR_Variation}
	\end{figure}

	Similarly in Fig.~\ref{fig:RMSE_Simulations_SnapshotsVarying}(b), we considered the APS ULA with an SNR of 10 dB. For the APS ULA, only formulation in~\eqref{eq:AVS_Msensor_Nsource_multi_snapshot_scenario_sparse_model_opt_problem_convex} is considered and the results of the proposed methodology are compared with the Paulraj-Kailath~\cite{paulraj1985direction} approach. The RMSE of the DOA estimates of the proposed methodology follows same trend as seen for the AVS ULA in Fig.~\ref{fig:RMSE_Simulations_SnapshotsVarying}(a). On the other hand, although the calibration approach in~\cite{paulraj1985direction} achieves the optimal solution, it requires more snapshots to achieve similar performance as the proposed methodology.
	\paragraph*{Fixed number of snapshots and varying SNR}
	The variation of the RMSE in the DOA estimates with respect to a change in SNR for a fixed number of snapshots is considered in Fig.~\ref{fig:RMSE_Simulations_SNR_Variation}. The same setup as in Fig.~\ref{fig:RMSE_Simulations_SnapshotsVarying} is considered with $N = 3$ ($\boldsymbol{\theta} = [78^\circ, 90^\circ, 102^\circ]$) where the RMSE of the source at $90^{\circ}$ is presented. 
	In Figs.~\ref{fig:RMSE_Simulations_SNR_Variation}(a)~and~(b), we consider the AVS and the APS ULA, respectively, with 1000 snapshots and varying SNR. Similar to Fig.~\ref{fig:RMSE_Simulations_SnapshotsVarying}, it is seen that 
	after calibration using the formulation in~\eqref{eq:DetModel} as well as in~\eqref{eq:AVS_Msensor_Nsource_multi_snapshot_scenario_sparse_model_opt_problem_convex} the RMSE of the DOA estimates decreases as the SNR increases for both the $l_1$ based spectra and the MUSIC spectra. Also as expected we can observe that the MUSIC spectra based DOA estimates outperform the $l_1$ based DOA estimates for a given SNR. Further, it can be inferred that the RMSE of the DOA estimates based on the proposed element-space model calibration technique asymptotically approaches the ideal scenario with no calibration errors. On the other hand, we can observe that the RMSE in the DOA estimates using the Weiss-Friedlander approach in Fig.~\ref{fig:RMSE_Simulations_SNR_Variation}~(a) for the AVS ULA and the Paulraj-Kailath approach in Fig.~\ref{fig:RMSE_Simulations_SNR_Variation}~(b) for the APS ULA, initially decreases as the SNR increases. However for an SNR greater than 5 dB the RMSE of the DOA estimates tends to saturate due to the finite sample errors in the covariance matrix estimation.

	\section{Experimental Results} \label{sec:ExperimentalResults}
	An experimental study was conducted in order to demonstrate 
	the proposed joint DOA and calibration algorithm for AVS arrays. As discussed, each AVS consists of a pressure microphone
	and 
	several orthogonal particle velocity transducers. A particle velocity transducer is commonly referred to as a Microflown~\cite{de2007microflown}. 
	A reliable calibration procedure is crucial for relating the sensor output to the physical quantity perceived.
	Unlike microphone calibration, there are no standardized procedures yet defined for characterizing
	the broadband response of particle velocity sensors.
	
	Microflown sensors were originally calibrated using a sound pressure microphone as a reference in a standing wave tube~\cite{bree1999realisation}, where the ratio between sound pressure and particle velocity (i.e., acoustic impedance) is well understood. Novel methods were later proposed for covering a wider frequency range, such as the "Piston-On-a-Sphere" technique (POS)~\cite{jacobsen2006note}. This approach relies on a sound source of known impedance measured in free field conditions and it achieves good results at mid and high frequencies. Thereafter, the POS technique was extended to lower frequencies by also measuring the acoustic pressure inside the sound source~\cite{basten2010full}. As a result, a full-bandwidth calibration procedure is now available by combining two measurement steps. In this section, the DOA estimation results based on the calibrated data using the POS technique (referred to as POS calibration), the Weiss-Friedlander approach~\cite{weiss1990eigenstructure} and the proposed calibration techniques (both the element-space and co-array approaches) are presented. 
	
\begin{figure}[h!]
	\centering
	\includegraphics[width=0.7\linewidth]{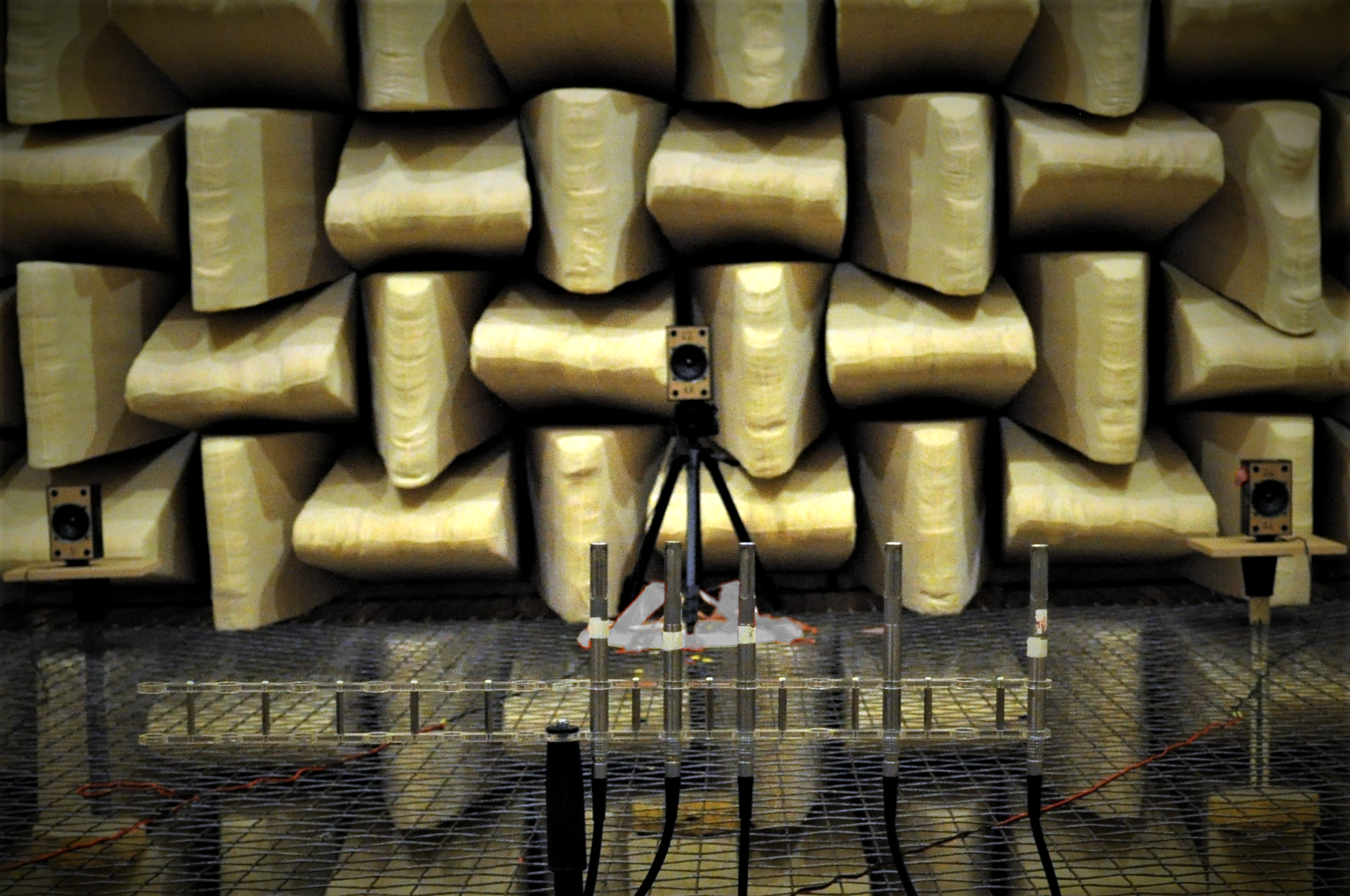}\\
	\caption{Picture of the experimental setup considering five AVSs and three speakers, located at a radius of $r = 3.6$ m.}
	\label{fig:ExperimentalSetup}
	\end{figure}

	A picture of the experimental setup is shown in Fig.~\ref{fig:ExperimentalSetup}, where five AVSs are seen arranged in a linear array configuration along with three speakers. 	
	The smallest inter-sensor spacing was $d = 0.05$ m with sensors located at positions ${\bf p}= [0, 1, 2, 4, 6]^{T}$ and the speakers were located along the circumference of a circle of radius $r = 3.6$ m with respect to the reference AVS in the array (the distance to the sources is more than 20 times the aperture of the array and therefore satisfying the far-field condition). The measurements were carried out in a fully anechoic chamber of the Faculty of Applied Physics of TU Delft (Netherlands) using uncorrelated white Gaussian excitations driving multiple 3 inch loudspeakers (resulting in high SNRs of approximately 30 dB). An Heim DATaRec 24 channels acquisition device with a sampling frequency of 25 kHz was used to record the data. The acoustic pressure and particle velocity information at a given frequency were obtained by computing a short time Fourier transform (STFT). Each recording was fragmented into segments of 1024 samples with $50\%$ overlap. A Hanning window was applied to each data segment prior to the STFT.		
	
\begin{figure}[h!]
		\begin{minipage}[h]{1\linewidth}
			\centering
			\includegraphics[width=0.8\linewidth]{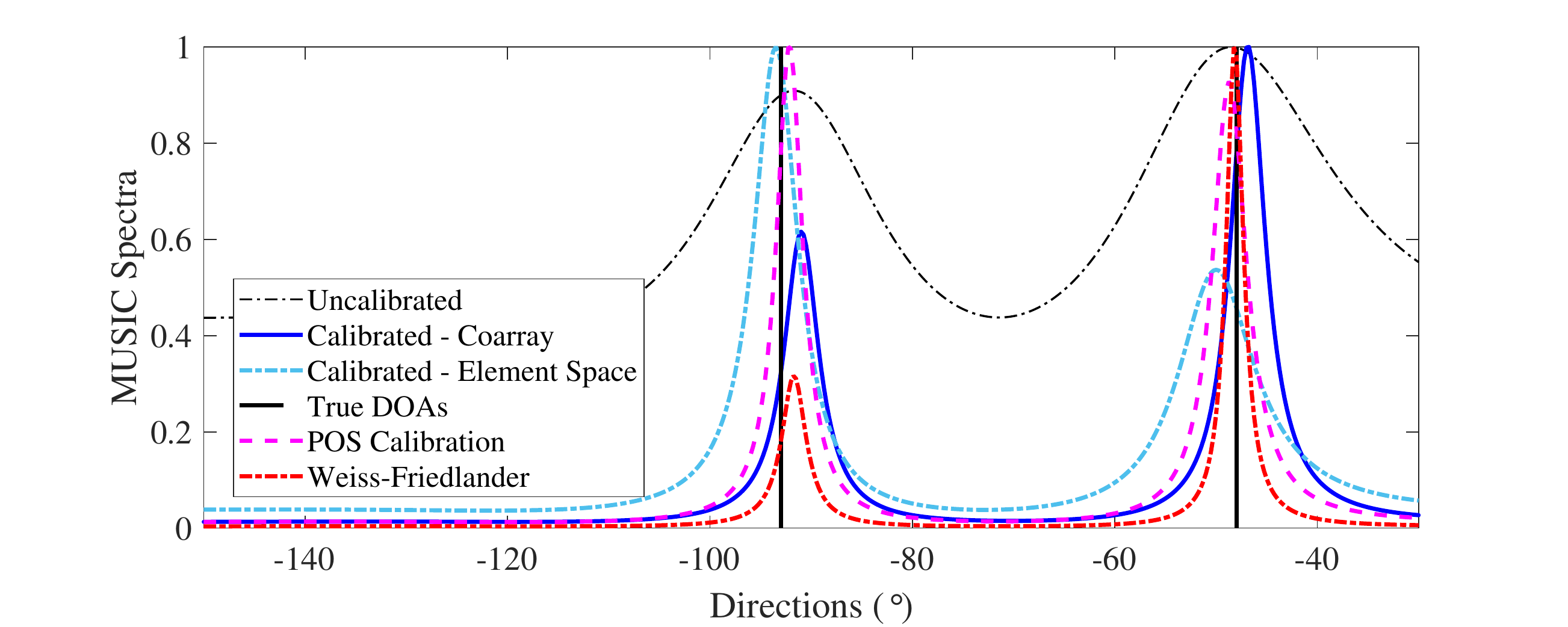}\\
			{\footnotesize(a) N = 2, $\boldsymbol{\theta} = [-45^{\circ}, -90^{\circ}]^{T}$.}
		\end{minipage} \hfill  
		\begin{minipage}[h]{1\linewidth}
			\centering
			\includegraphics[width=0.8\linewidth]{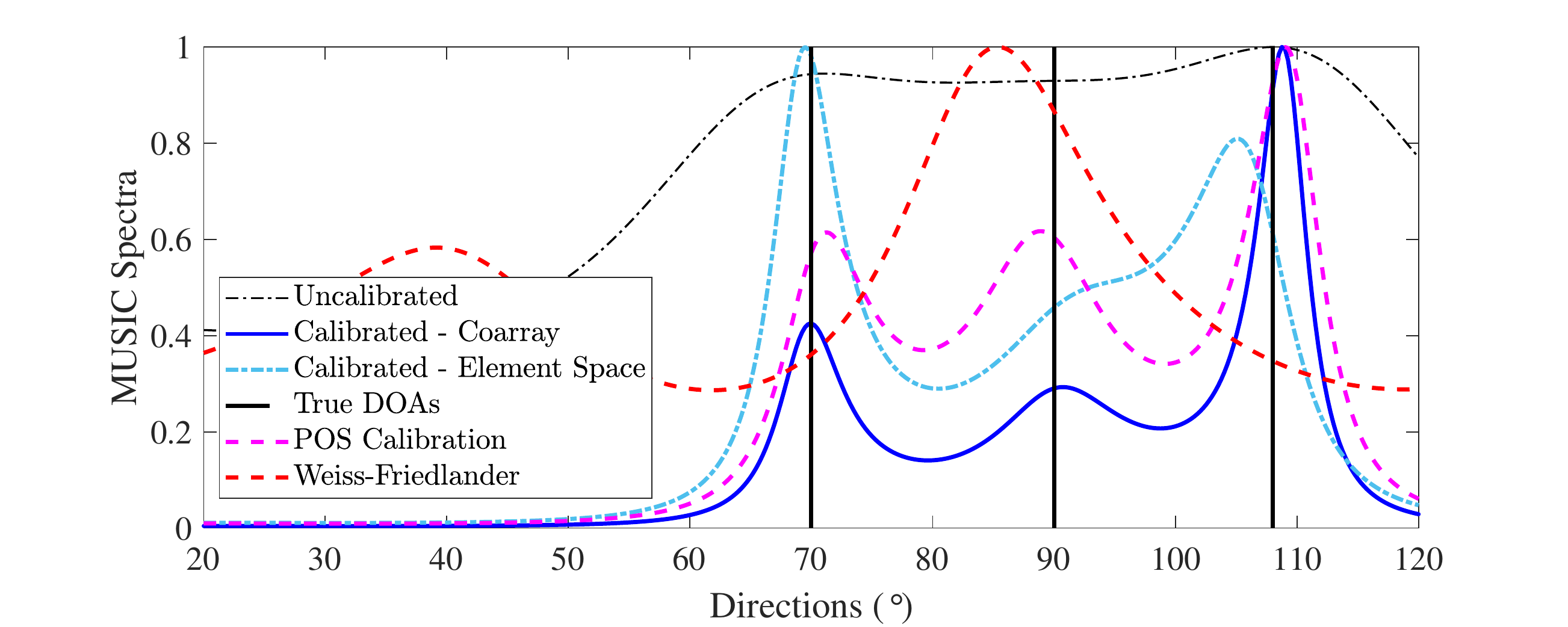}\\
			{\footnotesize{(c) N = 3, $\boldsymbol{\theta} = [70^{\circ}, 90^{\circ}, 108^{\circ}]^{T}$.}}
		\end{minipage}
		\caption{MUSIC spectra based DOA estimates using an AVS array with $M = 5$, $N = 2$ and $f = 2000$ Hz. The true DOAs are indicated by the black solid lines.}
		\label{fig:ExperimentalResults}
	\end{figure}

	The raw output signals from all the five AVSs at a time instant $t$ for a particular frequency bin were collected in a vector ${\bf x}(t)$, similar to~\eqref{eq:data_model}. Without loss of generality, we have considered the first channel of the first AVS in the array as the reference channel with known gain and phase response which is sufficient to obtain a unique solution as seen in the identifiability conditions for AVS arrays. The joint DOA and calibration algorithm based on~\eqref{eq:DetModel}~and~\eqref{eq:AVS_Msensor_Nsource_multi_snapshot_scenario_sparse_model_opt_problem_convex} were applied on the captured measurement data ${\bf x}(t)$ consisting of $L = 1000$ snapshots at a frequency of $f = 2000$ Hz. The corresponding grid-free MUSIC spectra based on the post-calibration measurement data are presented in Fig.~\ref{fig:ExperimentalResults}.
	
	In Fig.~\ref{fig:ExperimentalResults}~(a)~and~(b), we considered two of the three speakers with $\boldsymbol{\theta} = [-45^{\circ}, -90^{\circ}]^{T}$ and three speakers that are closely spaced with $\boldsymbol{\theta} =[70^{\circ}, 90^{\circ}, 108^{\circ}]^{T}$, respectively. 	
%
We can observe that for the uncalibrated data, the resolution of MUSIC is poor. However, improved spectra with higher resolution can be seen after compensating with the estimated calibration parameters. The MUSIC spectrum obtained from~\eqref{eq:AVS_Msensor_Nsource_multi_snapshot_scenario_sparse_model_opt_problem_convex}, results in a high resolution comparable to the results that are obtained with the reference POS calibration approach. However, the spectrum obtained from~\eqref{eq:DetModel}, has a lower resolution (especially in the three source case) and shows a small bias compared to the co-array domain based solver. The, Weiss-Friedlander approach results in degraded estimates compared to the proposed approach, specifically in Fig.\ref{fig:ExperimentalResults}~(b) it can be observed that none of the sources are resolved.

	\section{Concluding Remarks}
	In this paper, we proposed a self calibration technique for both the element-space and co-array data models that is applicable to both acoustic pressure and vector sensor arrays. Also, we derived and discussed a number of identifiability conditions for all the considered cases under which a unique solution for both the calibration parameters and the source DOAs can be obtained. It is interesting to note that for the AVS array, irrespective of the considered geometry, it is possible to calibrate all the sensors with respect to only one of the channels in the array.	
	
	Based on the proposed approach, we showed that it is indeed possible to jointly estimate calibration errors and source directions using a \emph{one-step} approach by exploiting the underlying algebraic structure and convex optimization techniques. It is shown that for infinite data records, we can in fact obtain the optimal solution suggesting the feasibility of the convex relaxations for both the element-space and co-array data models. However, when the number of time snapshots are limited and we have a pre-defined grid, we stated that the proposed methodology can be used as a pre-conditioning step to estimate the calibration errors. Then a grid-free approach such as MUSIC/SS-MUSIC can be applied on the gain and phase errors compensated measurement data to obtain improved and reliable DOA estimates. Furthermore, through simulations, we showed that even for finite data records we are able to recover all the source DOAs and we perform better than the existing calibration techniques for all the considered scenarios. Finally, experimental results based on real measurement data with an AVS linear array that are collected in an anechoic chamber are presented to showcase the effectiveness of the proposed calibration techniques using both the element-space and co-array data model.

	\bibliographystyle{IEEEtran}
	\bibliography{tspCalibration_Rev4}
	
\end{document}